\documentclass[a4paper,11pt]{article}
\pdfoutput=1 % if your are submitting a pdflatex (i.e. if you have
\usepackage{jcappub} % for details on the use of the package, please
\usepackage[T1]{fontenc} % if needed
\usepackage{subcaption}

\title{\boldmath Fast Forward Modelling of Galaxy Spatial and Statistical Distributions}

\author[a,1]{Pascale Berner, \note{Corresponding author.}}
\author[a]{Alexandre Refregier,}
\author[a]{Beatrice Moser,}
\author[a, b]{Luca Tortorelli, }
\author[a]{Luis Fernando Machado Poletti Valle}
\author[a, c]{and Tomasz Kacprzak}

\affiliation[a]{Institute for Particle Physics and Astrophysics, ETH Zurich, Wolfgang-Pauli-Str. 27, 8093 Zurich, Switzerland}
\affiliation[b]{University Observatory, Faculty of Physics, Ludwig-Maximilian-Universität M\"unchen, \\ Scheinerstrasse 1, 81679 Munich, Germany}
\affiliation[c]{Swiss Data Science Center, Paul Scherrer Institute, Forschungsstrasse 111, 5232 Villigen, Switzerland}

\emailAdd{pascale.berner@phys.ethz.ch}
\emailAdd{alexandre.refregier@phys.ethz.ch}
\emailAdd{moserb@phys.ethz.ch}
\emailAdd{luca.tortorelli@physik.lmu.de}
\emailAdd{lmachado@phys.ethz.ch}
\emailAdd{tomaszk@phys.ethz.ch}

\abstract{
A forward modelling approach provides simple, fast and realistic simulations of galaxy surveys, without a complex underlying model. For this purpose, galaxy clustering needs to be simulated accurately, both for the usage of clustering as its own probe and to control systematics. We present a forward model to simulate galaxy surveys, where we extend the Ultra-Fast Image Generator to include galaxy clustering. We use the distribution functions of the galaxy properties, derived from a forward model adjusted to observations. This population model jointly describes the luminosity functions, sizes, ellipticities, SEDs and apparent magnitudes. To simulate the positions of galaxies, we then use a two-parameter relation between galaxies and halos with Subhalo Abundance Matching (SHAM). We simulate the halos and subhalos using the fast PINOCCHIO code, and a method to extract the surviving subhalos from the merger history. Our simulations contain a red and a blue galaxy population, for which we build a SHAM model based on star formation quenching. For central galaxies, mass quenching is controlled with the parameter M$_{\mathrm{limit}}$, with blue galaxies residing in smaller halos. For satellite galaxies, environmental quenching is implemented with the parameter t$_{\mathrm{quench}}$, where blue galaxies occupy only recently merged subhalos. We build and test our model by comparing to imaging data from the Dark Energy Survey Year 1. To ensure completeness in our simulations, we consider the brightest galaxies with $i<20$. We find statistical agreement between our simulations and the data for two-point correlation functions on medium to large scales. Our model provides constraints on the two SHAM parameters M$_{\mathrm{limit}}$ and t$_{\mathrm{quench}}$ and offers great prospects for the quick generation of galaxy mock catalogues, optimized to agree with observations.
}

\begin{document}
\maketitle
\flushbottom

\section{Introduction}
\label{sec:intro}
One of the main focuses of modern cosmology is the study of the distribution of matter in the universe. We know that most of the matter in the universe is composed of dark matter, which can only be observed indirectly, for example with gravitational lensing. Visible matter on the other hand is in the form of gas and galaxies and can be observed in different ranges of the electromagnetic spectrum.\\
Galaxy surveys are one of the most important ways to probe the universe. Galaxy shapes, distortions and magnifications allow for weak lensing analyses (e.g. Dark Energy Survey (DES, \cite{Abbott_2018}), Kilo-Degree Survey (KiDS, \cite{kids2013})). The clustering of galaxies is a powerful cosmological probe and also a systematic effect for lensing analyses.\\
To understand and evaluate the observed galaxy clustering we rely on simulations. They are important for example to estimate uncertainties of observational data, to build analysis pipelines, and for cosmological parameter estimation. For a simulation to be useful, it must be realistic and accurate enough and yet cover a large enough volume. This becomes a trade-off for the finite computational resources.\\
Simulations that contain only dark matter are simpler, since they do not require a treatment of the baryons and therefore allow larger numbers of simulation particles. These N-body simulations describe the motion of the particles, often through Newtonian gravity, thereby predicting structure formation (e.g. \texttt{GADGET-2} \cite{gadget2001, gadget2005}, \texttt{GADGET-4} \cite{gadget4_2021}, \texttt{PKDGRAV3} \cite{potter2016pkdgrav3} and \texttt{ABACUS} \cite{abacus_2019}). A way to extract single objects from the resulting cosmic web is to use dark matter halos. A halo is a bound overdense region consisting of multiple simulation particles, describing objects, like the dark matter halo surrounding our Milky Way. Substructures are then described as subhalos, i.e., collapsed halos within another halo. Baryonic matter can be simulated using hydrodynamical simulations, but these are computationally expensive and therefore usually applied to smaller simulation volumes. Furthermore, the modelling of baryons is less well understood since the underlying theory is more complex than with dark matter only. \\
An alternative is to simulate the dark matter distribution and relate it to the distribution of baryons. This is possible for example for neutral hydrogen mass using a HI to halo mass relation (e.g. \cite{Padmanabhan_2017}) or for galaxies hosted by halos (e.g. \cite{2010_book_mo}). The latter requires an understanding of the relation between the dark matter and the galaxy distribution, or the relation between the galaxy properties and the properties of the dark matter halos they reside in. Since the field studying galaxy formation and evolution is still evolving and cannot easily be simplified, the galaxy to halo relation is an approximate model, which can nonetheless generate realistic galaxy distributions \cite{2010_book_mo}. \\
Commonly used methods for simulations to populate dark matter halos with galaxies are the Halo Occupation Distribution HOD (e.g. \cite{Berlind_2002, hod_2005, Jing_1998, Peacock_2000, Seljak_2000, Scoccimarro_2002, Zheng_2009, Hadzhiyska_2020}), (Sub-)Halo Abundance Matching (S)HAM (e.g. \cite{sham_2012, sham_2013, Kravtsov_2004, Tasitsiomi_2004, Vale_2004, Conroy_2006, Guo_2010, Trujillo_Gomez_2011, Reddick_2013}) and semi-analytic models (e.g. \texttt{GAEA} \cite{2004gaea, 2014gaea, 2016gaea, 2017gaea}). While semi-analytic models are powerful for many applications, they require treatment of the formation and clustering of galaxies and are thus more complicated. For HOD, one needs to specify the functions N$_{\mathrm{cen}}$ and N$_{\mathrm{sat}}$, standing for the average number of central and satellite galaxies respectively residing in a halo of a given mass at a given redshift. Since Halo Abundance Matching assigns only one galaxy to each halo or subhalo and derives a monotonic relation between halo and galaxy properties, SHAM does not require these two functions. Additionally, SHAM has the advantage of giving positions and velocities for all the galaxies, since every galaxy is assigned to a separate halo.\\
In this work, we simulate the distribution of galaxies with a forward model consisting of two steps. In the first step, the distribution functions of the galaxy properties were forward modelled and constrained with Approximate Bayesian Computation (ABC) \cite{Herbel_2017, Tortorelli_2020}, comparing to data using the image simulations by the Ultra-Fast Image Generator \texttt{UFig} \cite{ufig2013, Kacprzak_2020}. For this purpose, the whole galaxy population is split into two categories: Red galaxies are more quiescent and blue galaxies have more star formation. In this model, the luminosity functions, sizes, ellipticities, SEDs and apparent magnitudes in observed filter bands were jointly constrained.\\
As a second step, we are left to derive the relation between halo and subhalo mass and galaxy luminosity, using a SHAM approach for the red and blue galaxies. The dark matter simulation we use was described in \cite{2021_berner}. It allows us to efficiently simulate halos and subhalos at high resolution in a lightcone and is based on \texttt{PINOCCHIO} \cite{pinocchio2002, pinocchio2013, pinocchio2017}. We describe the halo-subhalo simulation in more detail in section \ref{sec:method_sub}. The (sub)halo mass we use for the SHAM is the one at the observed redshift for halos and the one at accretion time for subhalos. It has been shown by e.g. \cite{Wechsler_2018} that the subhalo mass at accretion is a good property to be used for matching galaxies to halos. The galaxy to halo relation could also be derived by using other halo properties (for example the highest mass in its history, the maximum circular velocity at observation, the peak maximum circular velocity in its history, or a combination of mass and virial velocity) and other galaxy properties (for example the galaxy size or the stellar mass) \cite{Wechsler_2018}. The dark matter simulation from \cite{2021_berner} does not have information on velocities within the halos, though, and our model to draw galaxies is based on luminosity functions.\\
Previous work has been done in this field. The project \texttt{PTHalos} \cite{Scoccimarro_2002} used halo occupation numbers to simulate the galaxy spatial distribution. The ELUCID project \cite{Yang_2018, Chen_2019} derives the galaxy-subhalo connection using a neighbourhood abundance matching method. The \texttt{UniverseMachine} \cite{Behroozi_2019} derives the star formation rate of galaxies from different halo properties, constrained by observed galaxy properties. \texttt{ScamPy} \cite{Ronconi_2020} uses subhalo clustering and abundance matching (SCAM) to paint galaxies onto halos and subhalos. In \cite{Guo_2016} they compare HOD with SHAM as well as SCAM. Another extension of subhalo abundance matching using hydrodynamical simulations and recipes for star formation rate was presented by \cite{Contreras_2020}. Several SHAM models were developed by comparing to SDSS and BOSS/eBOSS \cite{sdss_2011, sdssIII, sdssIV, sdssV}, e.g. \cite{2016_favole, 2022_jiaxi}. Recent work using the first data of DESI (EDR, Early Data Release \cite{2023_desiedr}) includes projects on SHAM \cite{2023_desi_berti, 2023_desi_gao, 2023_desi_jiaxi} and on HOD \cite{2023_desi_rocher, 2023_desi_sandy}.\\
Our goal is to develop a forward model that can be run many times due to its low computational cost. A discussion on forward modelling can be seen in e.g. \cite{Fagioli_2020}. The idea in forward modelling is to generate the same data format as provided by observations, evaluate both consistently for comparison, and then loop back to adapt the model parameters. Contrary to a semi-analytical model or a hydrodynamical simulation, our forward model is simple and in particular involves only two free SHAM parameters. In order to be able to use Approximate Bayesian Computation (ABC, e.g. \cite{2013weyant, Akeret_2015, Herbel_2017, Tortorelli_2020}) to optimize the galaxy model in the future, the galaxy to halo assignment needs to be very fast. Only two additional free parameters are added to the existing galaxy model of \texttt{UFig} \cite{ufig2013, Kacprzak_2020}. We also use the approximate dark matter halo catalogues described in \cite{2021_berner} instead of a full N-body simulation with a halo finder to speed up our method, since the underlying simulation for the dark matter halos and subhalos is by far the most expensive part of the pipeline.\\
This paper is structured as follows. In section \ref{sec:method}, we describe our procedure for assigning simulated galaxies to halos and subhalos. We describe the galaxy survey that we use to compare to our simulated galaxy catalogue in section \ref{sec:survey}. In section \ref{sec:results}, we show the results of the statistical properties and of the clustering of our simulated galaxy catalogue. Our conclusions are summarised in section \ref{sec:conc}.

\section{Method}
\label{sec:method}
In this section, we describe our approach to simulate galaxies including their spatial distribution. Our approach includes both the simulation of dark matter halos and subhalos and the subsequent assignment of galaxies to them.

\subsection{Halo and subhalo catalogue}
\label{sec:method_sub}
For the spatial distribution of galaxies, we use the simulated distribution of dark matter halos and subhalos by applying Subhalo Abundance Matching. The simulation of faint galaxies with accurate clustering requires populating both halos and subhalos with galaxies. This helps reproduce the fact that galaxies are also associated with subhalos, especially in high density regions such as galaxy clusters.\\
More distant galaxies are observed at an earlier stage of the universe. We thus need lightcones of halos and subhalos in order to populate them with galaxies. Since one of our main goals is to create fast simulations, we cannot depend on full N-body simulations with halo finders. They are computationally too expensive, especially since we need simulations with a high enough resolution for halos associated to single galaxies. We therefore use the halo-subhalo clustering simulations presented in \cite{2021_berner}.\\
These simulations are based on \texttt{PINOCCHIO} \cite{pinocchio2002, pinocchio2013, pinocchio2017}, a dark matter simulation that efficiently produces halos in simulation boxes and lightcones. \texttt{PINOCCHIO} is much faster than an N-body simulation, as it uses Lagrangian Perturbation Theory and builds groups of particles while running. It can thus be run at high resolution, while still using a large enough simulation volume. However, it does not directly simulate the subhalo catalogues.\\
In a previous work in this series \cite{2021_berner}, the subhalo progenitors for each halo in a snapshot or lightcone were extracted from the merger history. The survival of subhalos is modelled using a fit for the subhalo merger time. Radial density profiles are used for the spatial distribution of the surviving subhalos within their hosts. A comparison with simulations using the N-body simulation \texttt{GADGET-2} \cite{gadget2001, gadget2005} and the halo finder \texttt{Rockstar} \cite{behroozi_2012, Behroozi_2019} showed that the resulting subhalo velocity function and the two-point correlation function of subhalos are realistic. The halo-subhalo simulations by \cite{2021_berner} were shown to be of the order $\sim700$ faster than a full N-body simulation.\\
We use the lightcones calculated directly within \texttt{PINOCCHIO}, combined with the subhalo code by \cite{2021_berner}. With \texttt{PINOCCHIO}, we can create high resolution halo simulations in large cosmological volumes. For this work, we simulate lightcones with a half opening angle of 60 $\deg$, resulting in one quarter of the sky. We simulate until redshift $z=0.75$, which is sufficient given our selection of relatively bright galaxies in this work. To reach the necessary resolution of halos corresponding to single galaxies and even dwarf galaxies, we use a box size of 500 h$^{-1}$Mpc and 2048$^3$ simulation particles, and include halos and subhalos with at least 10 particles. This results in a minimum (sub-)halo mass of $1.1 \cdot 10^{10}$ h$^{-1}$M$_{\odot}$. Since \texttt{PINOCCHIO} is not a fully numerical simulation, it is possible for us to use halos with very few particles, as described in \cite{2021_berner}. For the simulations, we use the same cosmological parameters as in \cite{2021_berner}, namely $\Omega_{\Lambda} = 0.73$, $\Omega_{m} = 0.27$, $h = 0.7$, $\Omega_{b} = 0.045$, $\sigma_8 = 0.811$, $n = 0.961$ and $w = -1$, which is close to the WMAP Cosmology \cite{2013_wmap}. To estimate the statistical uncertainty, we run multiple simulations with the same cosmology, the same resolution, the same lightcone settings but different initial conditions.

\subsection{Galaxy assignment}
\label{sec:gal_assign}

\subsubsection{UCat}
\label{sec:ucat}
\begin{figure}[t]
\centering 
\includegraphics[width=.90\textwidth,angle=0]{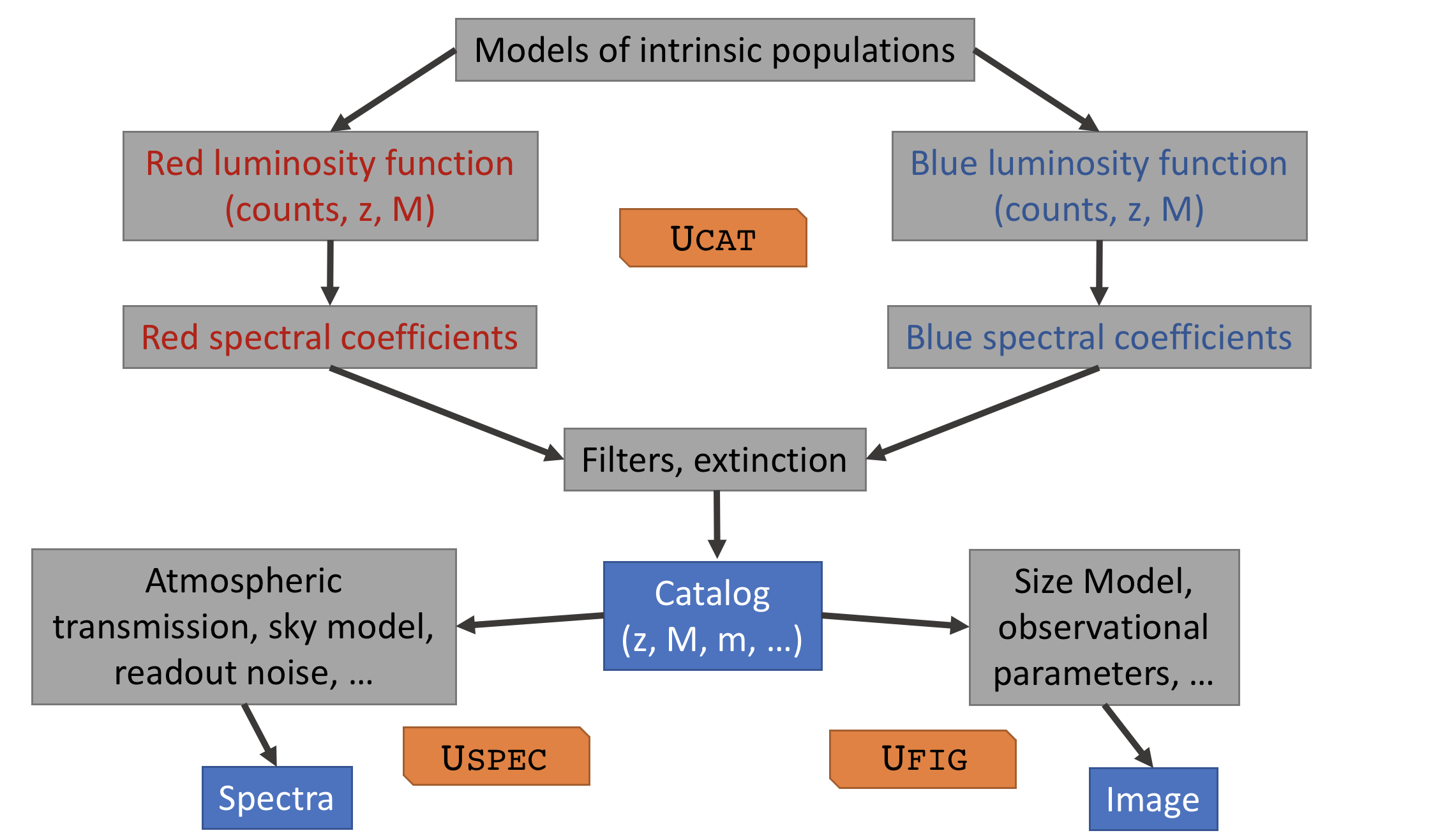}
\caption{Diagram showing the workings of \texttt{UCat}, \texttt{UFig} and \texttt{USpec}. \texttt{UCat} draws the galaxies on catalogue level without observational influence, \texttt{UFig} adds survey-specific properties to simulate realistic images and \texttt{USpec} adds a sky model and noise sources to simulate spectra.}
\label{fig:ucat_ufig}
\end{figure}

The simulated galaxies for the abundance matching and the final simulation are taken from the framework of the Ultra-Fast Image Generator (\texttt{UFig}, \cite{ufig2013, Bruderer_2016}). \texttt{UFig} simulates images for astronomical applications by getting a catalogue and rendering images that include observational and instrumental effects. Observational properties include spectral coefficients, while instrumental effects include the point spread function, noise and pixelization. Originally developed for the Monte-Carlo Control Loops Pipeline (\texttt{MCCL}, \cite{refregier2014, Kacprzak_2020}), \texttt{UFig} was designed for forward modelling and therefore to be fast.\\
The underlying models are simple, which beside code optimization is the reason for \texttt{UFig}'s speed. The catalogue generator for \texttt{UFig} is called \texttt{UCat} and the galaxy population model was developed by \cite{ufig2013, Herbel_2017}. The total galaxy population is split into a redder and a bluer population, with bluer galaxies having higher star formation rates. Whether a galaxy is red or blue in this simulation is an intrinsic property and not decided according to a colour cut, as is custom for galaxy surveys. \texttt{UCat} estimates the number of red and blue galaxies per volume and applies that to a given redshift range and field of view (i.e. lightcone aperture or image size). The code then samples red and blue galaxies separately from luminosity functions, which are expressed in absolute magnitudes M. The luminosity functions $\mathrm{\Phi(M,z)}$ as a function of absolute magnitude and redshift were optimized together with the rest of the model parameters of \texttt{UFig} in \cite{Herbel_2017} and \cite{Tortorelli_2020} using Approximate Bayesian Computation (ABC, \cite{Akeret_2015}). A more detailed description of the luminosity functions can be found in \cite{Tortorelli_2020}. We use the posterior from the ABC of \cite{Herbel_2017} for this work. The luminosity functions for red and blue galaxies are displayed in appendix \ref{sec:app_a}.\\
\texttt{UCat} then calculates apparent magnitudes in different filter bands from the absolute magnitudes and redshifts by using spectral coefficients, filter throughput and extinction values. The five spectral coefficients for each galaxy give a combination of the \texttt{kcorrect} spectral templates \cite{Blanton_2007_kcorrect}. The procedure is shown schematically in figure \ref{fig:ucat_ufig}. Note that additionally, the spectra simulator \texttt{USpec} \cite{Fagioli_2018, Fagioli_2020} can calculate realistic spectra from the spectral coefficients by adding a sky model, atmospheric transmission and readout noise.\\
Before this work, the simulated galaxy positions were randomly selected. In this work, we introduce a method to simulate galaxy catalogues with realistic clustering properties.

\subsubsection{Abundance matching}
\label{sec:matching}
\begin{figure}[t]
\begin{subfigure}{.5\textwidth}
  \centering
  % include first image
  \includegraphics[width=1.\linewidth]{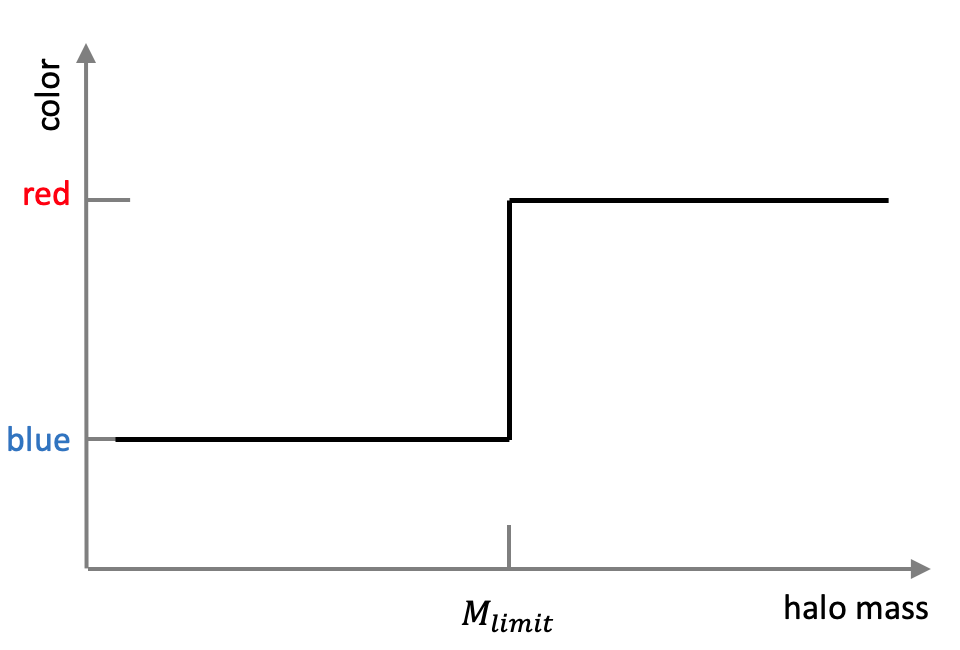}  
  \caption{Red vs. blue for halos}
  \label{fig:sham_halos}
\end{subfigure}
\begin{subfigure}{.5\textwidth}
  \centering
  % include second image
  \includegraphics[width=1.\linewidth]{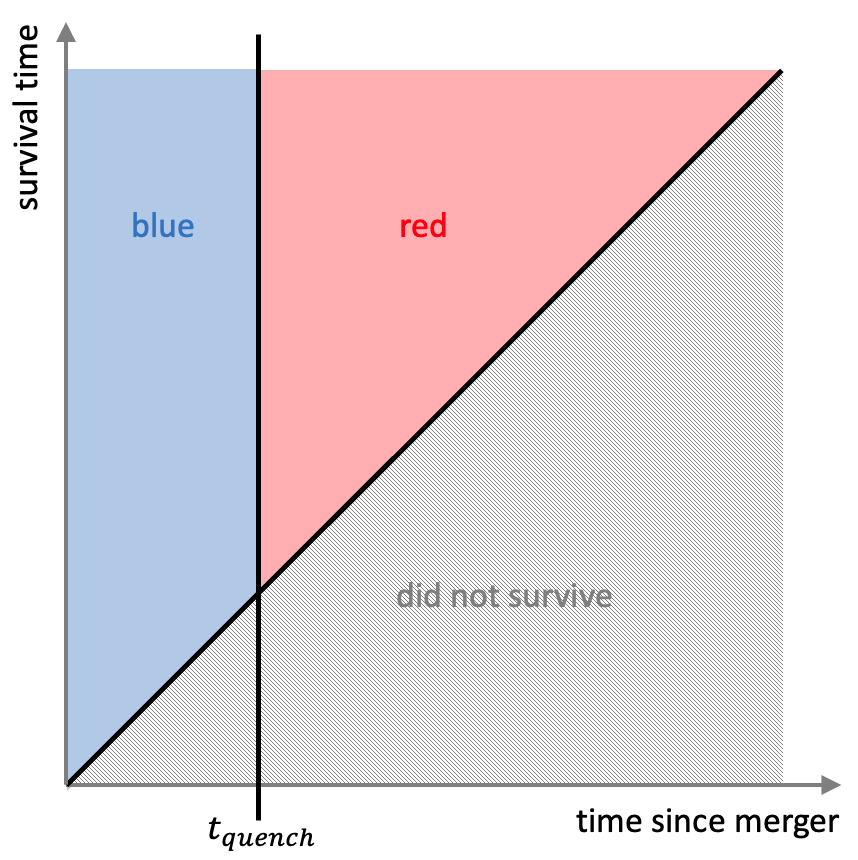}  
  \caption{Red vs. blue for subhalos}
  \label{fig:sham_subhalos}
\end{subfigure}
\caption{Diagrams depicting the SHAM model based on star formation quenching. Halos host blue central galaxies below a certain mass limit (left figure, \ref{fig:sham_halos}). Subhalos host blue satellite galaxies until a certain quenching time (right figure, \ref{fig:sham_subhalos}).}
\label{fig:red_blue_halo_sub}
\end{figure}

Abundance Matching derives a monotonic relation between (sub)halo and galaxy properties by matching the cumulative numbers of halos and galaxies per volume, assuming that each halo in the catalogue hosts one galaxy. In our case, the matching is done based on the assumption that more luminous galaxies are hosted by more massive halos. The number density of (sub)halos and galaxies, as well as the luminosity function of galaxies, depends on redshift. We therefore match the abundance in small bins of redshift. To ensure that also the simulated volume is the same, we match the field of view.\\
The clustering of galaxies does not only depend on the intrinsic luminosity, but also on the galaxy type. A simple way to split the whole population is to distinguish between red and blue galaxies. Red galaxies are passive, with little star formation, while blue galaxies have more star formation and are typically younger. The spatial distribution of halos depends on (sub)halo mass and on whether that halo is a halo or a subhalo. A galaxy assigned to a halo is called a central galaxy (especially if there are subhalos in this halo), while the galaxies assigned to subhalos are satellites. There has been plenty of research into the relation between red and blue and central and satellite galaxies (see e.g. \cite{Bell_2004, Willmer_2006, Cooper_2007, van_den_Bosch_2008, 2010_book_mo, 2011_zehavi, 2018_xu, Wechsler_2018, Girelli_2020}).\\
Red galaxies are typically older than blue galaxies, having already lost most of their active star formation due to being stripped of their cold gas. Galaxy clusters also have a higher abundance of red galaxies compared to the whole Universe. Furthermore, the central galaxy in a group or cluster, therefore the galaxy associated to a massive halo, is more often red than blue. Overall, blue galaxies on average reside in less dense regions than red galaxies (see e.g. \cite{Collister_2005, Elbaz_2007, Tang_2020}).\\
We therefore assign red galaxies to older subhalos and to halos above a certain halo mass M$_{\mathrm{limit}}$. The distinction between red and blue satellites is done using the satellite quenching time t$_{\mathrm{quench}}$, where a satellite is blue until it has been part of its host for longer than t$_{\mathrm{quench}}$. This procedure is shown in figure \ref{fig:red_blue_halo_sub}. A step function without any spread is a simplification, but our model is designed to be as simple as possible. The mass limit M$_{\mathrm{limit}}$ and the quenching time of satellites t$_{\mathrm{quench}}$ are free parameters of our model and can be determined to ensure the galaxy statistics are conserved. Subhalos that have been part of their host for longer than their expected survival time according to the survival model from \cite{2021_berner} are not included in our simulation, and therefore not assigned a galaxy.
\\
\noindent Each halo and subhalo in the lightcone is flagged to either host a red or a blue galaxy. The galaxies drawn from \texttt{UCat} also have a well-defined flag to be either red or blue. The relation between halo mass and galaxy luminosity is therefore done separately for red and blue galaxies.\\
We bin the halos in redshift and in logarithmic mass, to get a two-dimensional histogram. The number of bins, both in redshift and in $\log(\mathrm{M}_{\mathrm{halo}})$, has to be adjusted. Too few bins give a relation that is too approximated, which is especially problematic when too few mass bins are used. Too many bins lead to gaps and instability in the halo histograms. We found that a good bin width in redshift is $\Delta z = 0.01$. In mass, we're using 60 bins with logarithmic spacing between the minimal and maximal halo mass. After binning the halos and subhalos in mass and redshift, we bin the galaxies in redshift, with the same binning edges.\\
For each redshift bin, we then start at the highest mass bin and match the numbers with those of the most luminous galaxies and save the minimum and maximum absolute magnitude for this mass bin. Removing the matched galaxies, we proceed with the next lower mass bin and do the same, continuing until the lowest mass bin that contains halos.\\
When creating the mass to luminosity relation, it is important to have a large enough halo-subhalo catalogue with a high enough resolution. Otherwise, when applying the mass to luminosity relation to a catalogue for a different simulation, there may be halos or subhalos for which the corresponding absolute magnitude is not specified.
\begin{figure}[t]
\centering 
\includegraphics[width=.80\textwidth,angle=0]{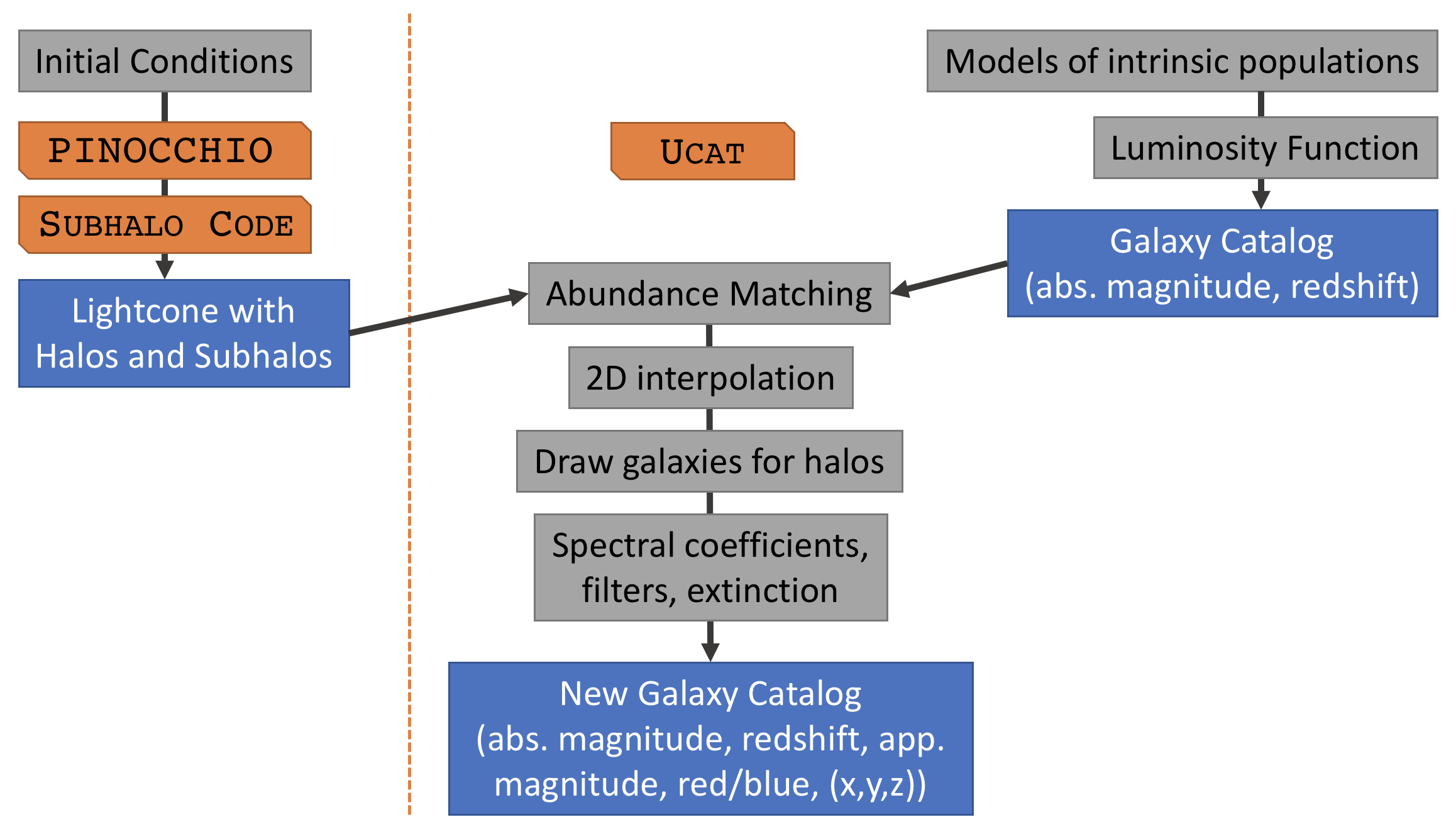}
\caption{Schematic description of the clustered galaxy simulations. The halo catalogue is created separately, while the interpolation and the drawing of galaxies is within \texttt{UCat}. The output in the end can be used for \texttt{UFig} the same way as in the version without clustering.}
\label{fig:ucat_new}
\end{figure} \\
\noindent In section \ref{sec:mass_lum_rel}, we show the derived relation between halo mass and absolute magnitude for red and blue galaxies. For each halo, we draw the absolute magnitude of its hosted galaxy using the derived two-dimensional interpolation in mass and redshift. The number of galaxies simulated this way is thus the same as the number of halos in the lightcone.\\
The halo redshift and the assigned absolute magnitude are then used to calculate the apparent magnitude with spectral coefficients, extinction values and filter throughputs within \texttt{UCat}. The filters bands have to be set according to the galaxy survey one wishes to simulate, and the filter throughputs are integrated once in \texttt{UFig} using the \texttt{kcorrect} templates and \texttt{K} corrections. Figure \ref{fig:ucat_new} shows a schematic of the procedure of creating the halo catalogue, deriving the interpolation, populating each halo with a galaxy and getting the apparent magnitudes.\\ \newpage
\noindent This halo and subhalo catalogue is complete in mass, therefore the new galaxy catalogue is complete in absolute magnitude. To simulate images with \texttt{UFig} or to simulate a galaxy survey in general, the catalogue can be cut by setting an upper limit for the apparent magnitude. It is thereby important to have a dark matter simulation with a high enough resolution, otherwise some of the faintest galaxies will be missing, especially at low redshifts.

\subsection{Magnitude shift calibration}
\label{sec:mag_cal}
\begin{figure}[b]
\centering 
\includegraphics[width=1.0\textwidth,angle=0]{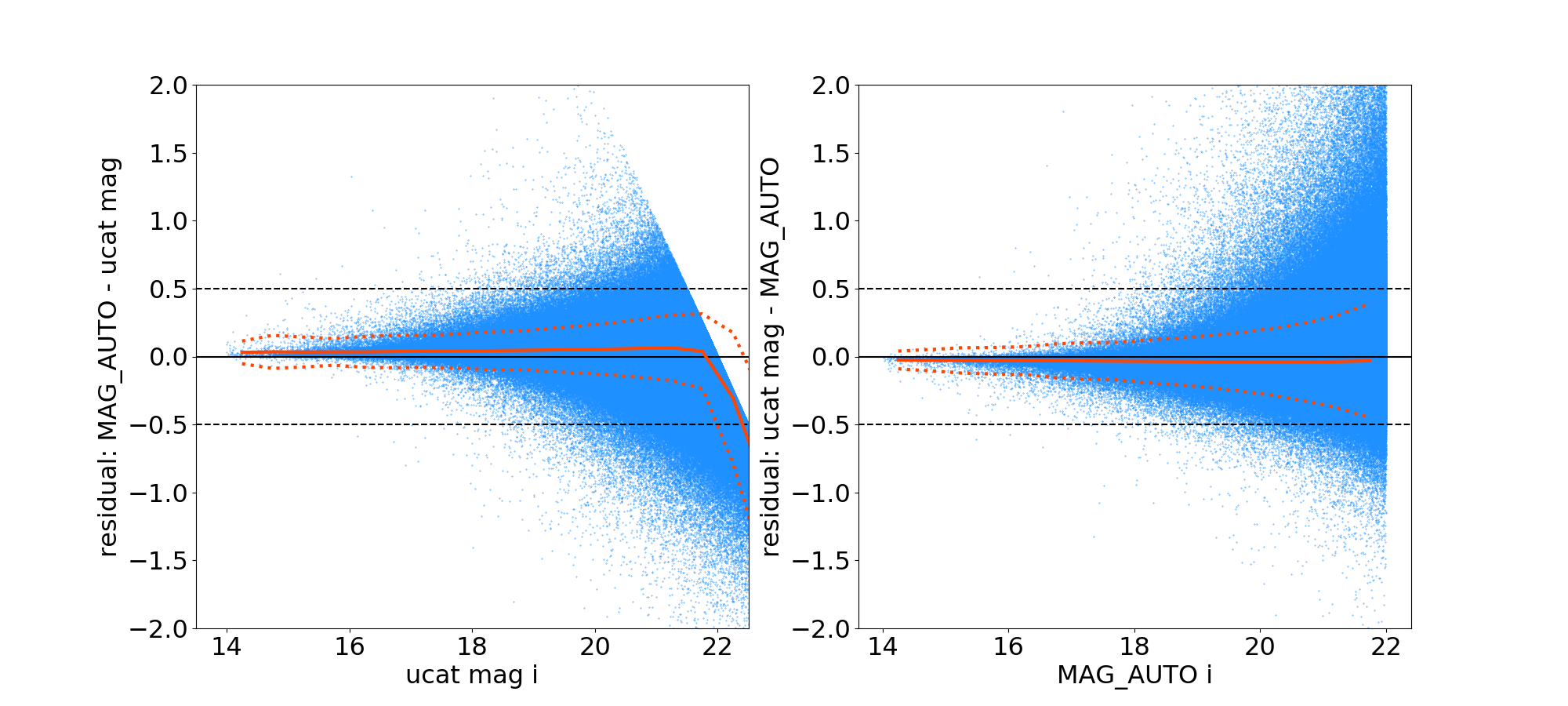}
\caption{Calibration of the shift in apparent magnitude $i$ with \texttt{Source Extractor}, using simulations. The solid orange line is the mean offset as a function of magnitude; the dotted lines are the 2$\sigma$ intervals. The calibration is done by simulating images using \texttt{UFig} and \texttt{Source Extractor} and then comparing the resulting magnitudes to the ones before creating images. On the left, the shift between real and \texttt{Source Extractor} magnitude is shown, on the right the opposite. We show dashed lines at $\pm$0.5 for reference.}
\label{fig:mag_shift}
\end{figure}

\noindent Our comparison to data is done at catalogue level. The data we compare to is collected by a telescope and in images, as described in section \ref{sec:survey}. The survey collaboration has run \texttt{Source Extractor} \cite{1996_sextractor} on their imaging data to produce catalogues. \texttt{Source Extractor} retrieves the sources from images and measures their fluxes and therefore their apparent magnitudes. Depending on the exact settings used in \texttt{Source Extractor}, the values for the apparent magnitudes are offset differently. Even small offsets in apparent magnitude that are not calibrated correctly will lead to shifted clustering results, especially when magnitude cuts or binning are involved.\\
The apparent magnitudes that we calculate using \texttt{UFig} for our catalogues do not have such an offset. \texttt{UFig} allows us to create realistic images. We therefore produced a small set of images, then ran \texttt{Source Extractor} with the same settings as the survey, and matched the retrieved objects to the ones in the original catalogue before the images. This allows us to calibrate the effect of \texttt{Source Extractor} on the apparent magnitudes in different filter bands. \\
For the i-band, the results of this calibration are shown in figure \ref{fig:mag_shift}. We have performed the same calibration also for the other magnitude bands. On the left, the shift from simulated magnitudes to observed magnitudes due to \texttt{Source Extractor} is shown. On the right, the opposite is shown, calibrating the shift to get from observed magnitude to true magnitude. The latter is useful for us, as we can either perform the correction by applying the first shift to the simulated catalogue or by applying the second shift to the data. Each blue dot corresponds to one galaxy, after removing stars. We calculate the mean and standard deviation of the offset in narrow bins of apparent magnitude since both shift and scatter are strongly magnitude dependent. The orange lines show the mean and 2$\sigma$ intervals of the offset. The black lines show -0.5, 0.0 and 0.5 shifts for clarity. For this calibration, we used objects reaching observed magnitudes of $i=22$, but our later analysis is restricted to brighter objects.\\
We apply the shift whenever comparing data to simulations. For figures \ref{fig:app_mag}, \ref{fig:scales_of_trust} and \ref{fig:2pcf}, we have shifted and added a scatter to the i-band apparent magnitude according to the calibration shown in figure \ref{fig:mag_shift}. For figure \ref{fig:color_space}, the shift of each magnitude band is applied, but not the scatter, since the scatter would have to be correctly correlated for the different bands for an analysis in colour space. As explained in more detail below, for figure \ref{fig:2pcf_redblue} we calibrated the magnitude shift separately for redder and bluer galaxies. For simplicity, and since applying the scatter has no noticeable effect on the measured clustering, we applied the shift to the data instead of the simulations for the chi-squared analysis presented in figure \ref{fig:chi2_surface}.

\section{Data}
\label{sec:survey}
To constrain the free model parameters M$_{\mathrm{limit}}$ and t$_{\mathrm{quench}}$ described in section \ref{sec:matching}, we compare our simulations to observational data. Unlike imaging surveys, spectroscopic surveys are usually not complete in a certain magnitude range. The Bright Galaxy Survey (BGS) of the Dark Energy Spectroscopic Instrument (DESI) \cite{2016desi} will provide a magnitude complete spectroscopic data set, but the survey is still running. Such incompleteness is complicated to forward model, as the same kind of objects have to be missing in the simulations as in the data, otherwise systematic errors are created. Furthermore, our model includes all galaxy types, not just a specific selection according to colour cuts. Since we do not want to build a model for just certain types of galaxies, we avoid any colour cuts, and therefore work with imaging data.\\
We compare our simulations to galaxies with $i<20$ from DES Y1, the public data release of the first year of observations of the Dark Energy Survey \cite{decam_2015, des_2016, Abbott_2018, 2018_des_y1, des_pipeline_2018, 2018_des_stargal, 2018_des_morph}. Both the data in catalogue form and the spatial, pixelized mask we use are publicly available from the DES Data Management.\footnote{https://des.ncsa.illinois.edu/releases/y1a1/gold}\\
For accurate forward modelling, we simulate the same cut-out of the sky as is covered by the data we use. The mask made publicly available by DES is in \texttt{HEALPix} \cite{2005_healpix} format with NSIDE=4096. For simplicity, since a smaller lightcone is easier to simulate, we reduce ourselves to the larger connected part of DES Y1 south of DEC < $-30 \deg$. This results in an area of 1685 $\deg^2$, after removing the pixels that are flagged with \textit{badmask}.\\
Our simulated galaxy catalogue is by construction free of stars and of any other objects besides galaxies. We therefore use the star-galaxy classification provided by DES and select only the objects classified as galaxies. Furthermore, we remove all the flagged objects, to ensure a clean data set. The corresponding columns are called \textit{MODEST\_CLASS} and \textit{FLAGS\_GOLD} in the data release. We note that \texttt{UFig} does simulate stars when rendering images, but we are only using the galaxy catalogues here.
\begin{figure}[]
\centering 
\includegraphics[width=.80\textwidth,angle=0]{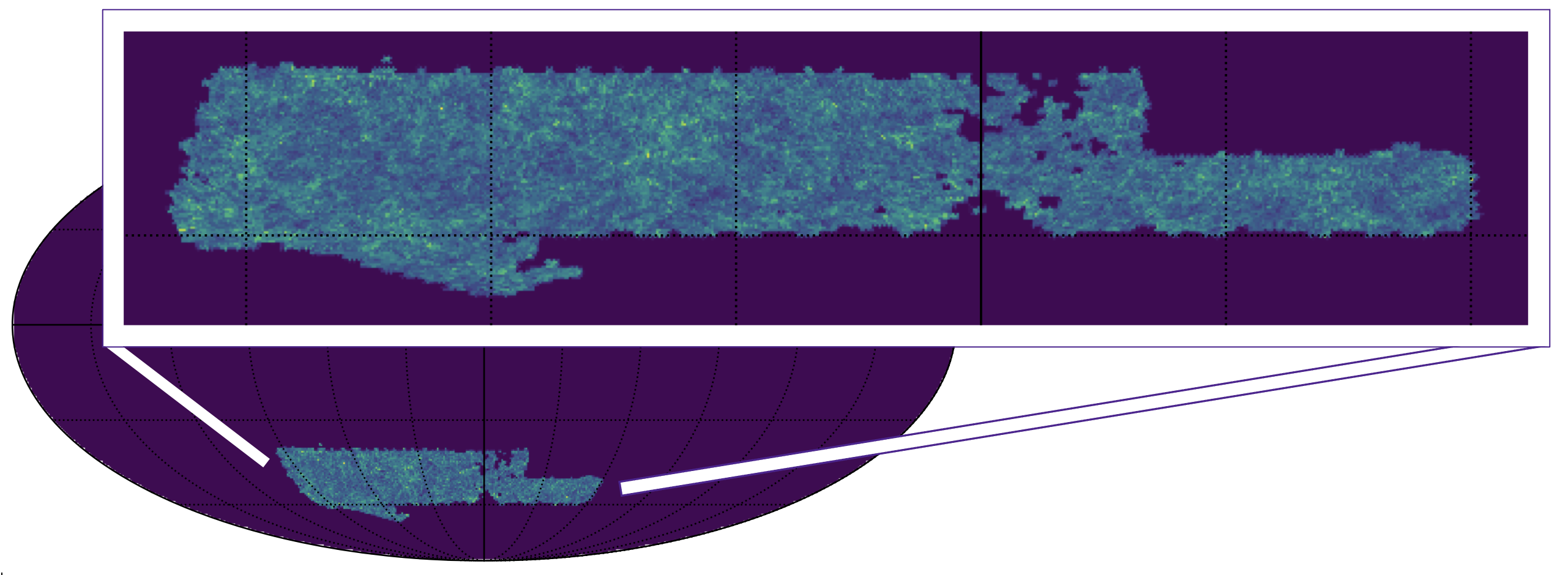}
\caption{Spatial mask used in this work, with the galaxies from DES Y1 with $i<20$. This figure uses NSIDE=256, while the mask used throughout this work has NSIDE=4096.}
\label{fig:mask}
\end{figure}\\
\noindent The area on the sky considered needs to be large enough to provide stable clustering information on small, medium, and large scales. Although an even larger area would increase the stability, the data provided by DES Y1 is enough to build our model. Furthermore, a wider survey would require a bigger lightcone, leading to more difficulties when simulating and processing it. Figure \ref{fig:mask} shows the spatial distribution of the galaxies from DES Y1 used in our comparison.\\
\noindent As described in section \ref{sec:matching}, we assign one galaxy to each halo and subhalo. At low redshifts, completeness in apparent magnitude requires a higher mass resolution for faint galaxies. Faint galaxies at low redshifts are small dwarf galaxies residing in very small halos and subhalos. Achieving the same completeness in the simulations as in the data is important for our forward model. Considering only the brightest galaxies of the dataset allows us to ensure that we do not miss galaxies in our simulations due to lack of resolution. We therefore apply a simple cut in apparent magnitude in the i-band, at $i=20$. At low redshifts, this data set still includes very small and faint galaxies.

\section{Results}
\label{sec:results}
In this section, we present our results derived from simulations and comparisons to the data described in the previous section. Our focus lies on the spatial distribution of galaxies and on the dependence of our simulations on the two free SHAM parameters.

\subsection{Galaxy distribution functions}
\label{sec:res_dist_func}

The galaxy population model in \texttt{UCat} and \texttt{UFig} is tuned via ABC such that the distribution of photometric properties of the galaxy sample agrees with observations (see \cite{Tortorelli_2020}). In appendix \ref{sec:app_a}, we show a comparison between the red and blue luminosity functions used here and observational results from other studies. In this work, the number of galaxies is given by the number of halos and subhalos. The abundance matching technique ensures the relation between halo mass and absolute magnitude, but does not maintain e.g., the number ratio between red and blue galaxies. We therefore perform some consistency checks of \texttt{UCat} before and after including clustering by using halos. We also compare the distribution of simulations and data.
\begin{figure}[]
\begin{subfigure}{.5\textwidth}
  \centering
  % include first image
  \includegraphics[width=1.\linewidth]{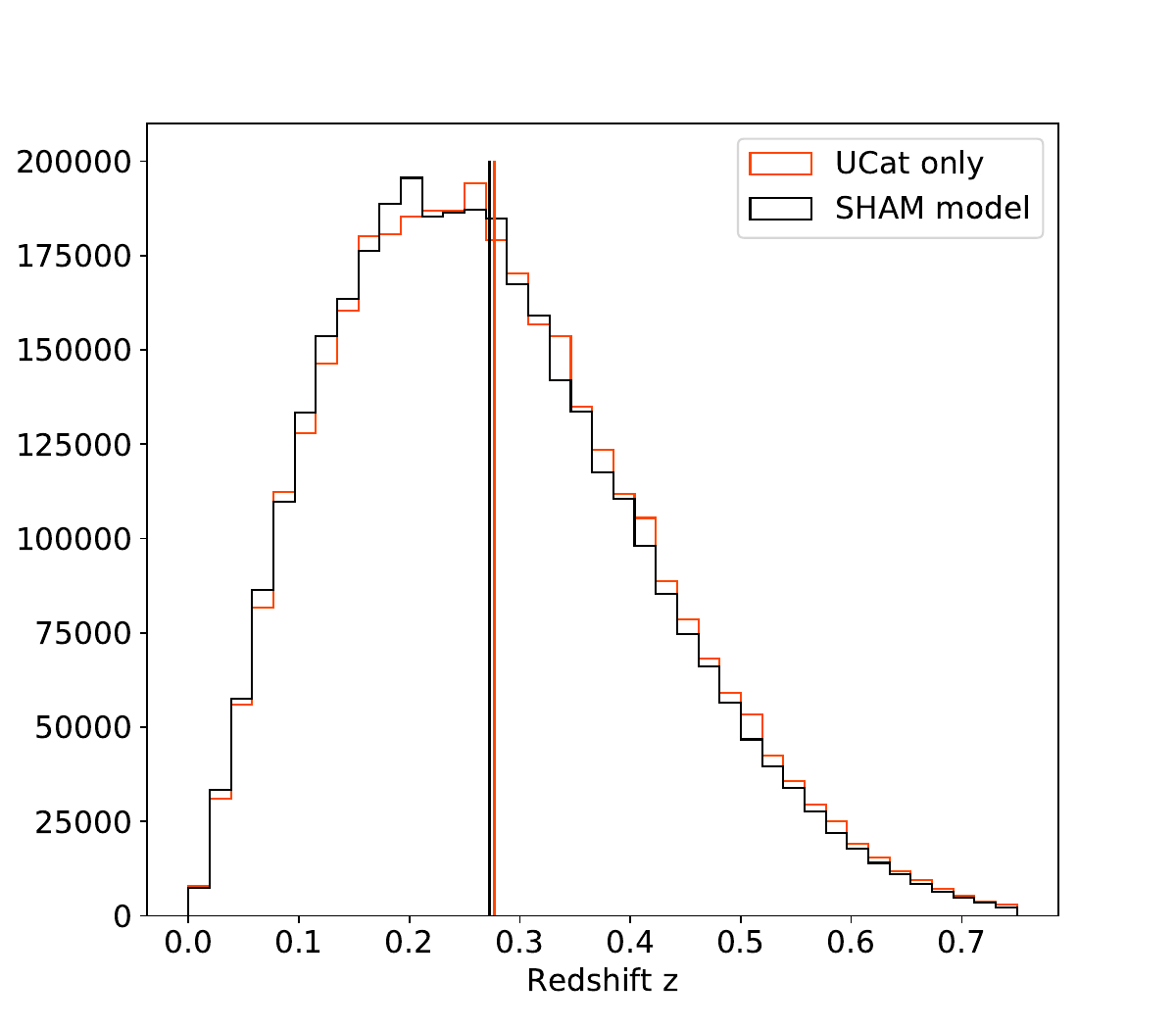}  
  \caption{Redshift distribution}
  \label{fig:n_z}
\end{subfigure}
\begin{subfigure}{.5\textwidth}
  \centering
  % include second image
  \includegraphics[width=1.\linewidth]{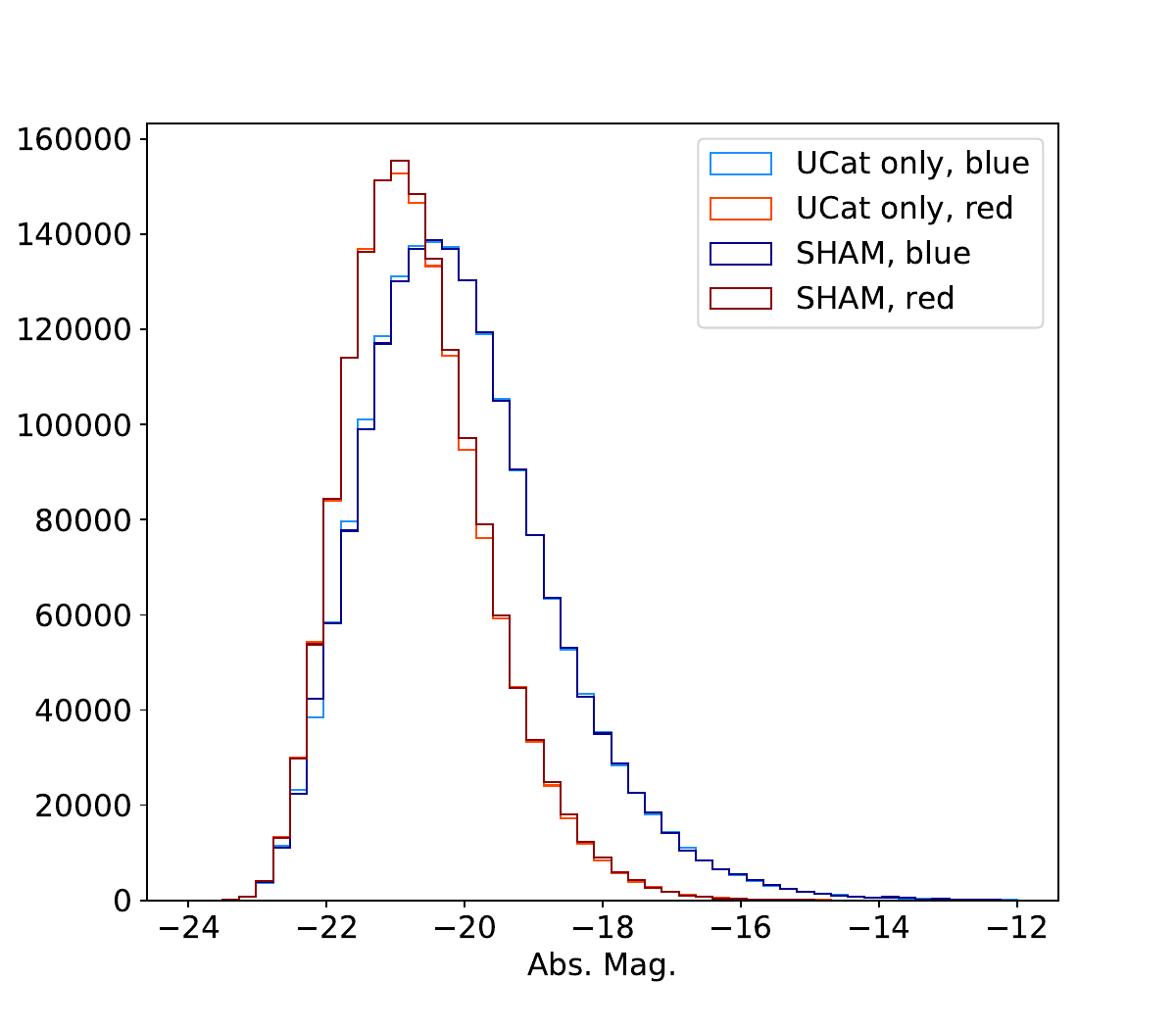}  
  \caption{Absolute magnitude distribution}
  \label{fig:lum_func}
\end{subfigure}
\caption{Histogram distributions of the simulated galaxies, comparing \texttt{UCat} only (light colours) and \texttt{UCat} with \texttt{PINOCCHIO} (dark colours). The redshift distribution is shown on the left with the vertical lines indicating the mean redshifts, the distribution of absolute magnitudes on the right.}
\label{fig:lum_func_n_z}
\end{figure}\\
\noindent Figure \ref{fig:n_z} shows the number of galaxies as a function of redshift, where the galaxies have an apparent magnitude below $i=20$. The orange line shows the distribution from \texttt{UCat} only, the black line shows it after including populating halos with galaxies.\\
By construction, the luminosity function of both red and blue galaxies is preserved. What may be changed is the minimum luminosity (maximum absolute magnitude) of the simulated galaxies. This can also be different for red and blue galaxies, if the number of halos set to host blue or red galaxies respectively is incorrect. Figure \ref{fig:lum_func} shows the distribution of absolute magnitudes for sampled red and blue galaxies when applying the cut at $i=20$, before and after using the abundance matching method. The consistency check presented in figure \ref{fig:lum_func_n_z} shows good agreement.\\
For forward modelling, the galaxy distributions of our simulations must agree with the data that we wish to simulate. The easiest way to check this is by looking at the histogram of apparent magnitudes. We do not have direct access to absolute magnitudes and estimating them for the data is difficult since we do not have accurate redshifts. Figure \ref{fig:app_mag} thus shows the histograms for the apparent magnitudes in the r and in the i band. The i-band apparent magnitude is sharply cut at $i=20$, by construction. We see that the simulation agrees well to the data. The luminosity function was fine-tuned in previous works involving \texttt{UFig} and \texttt{UCat}, and we use the posterior from the ABC presented in \cite{Herbel_2017}.
\begin{figure}[]
\centering 
\includegraphics[width=.990\textwidth,angle=0]{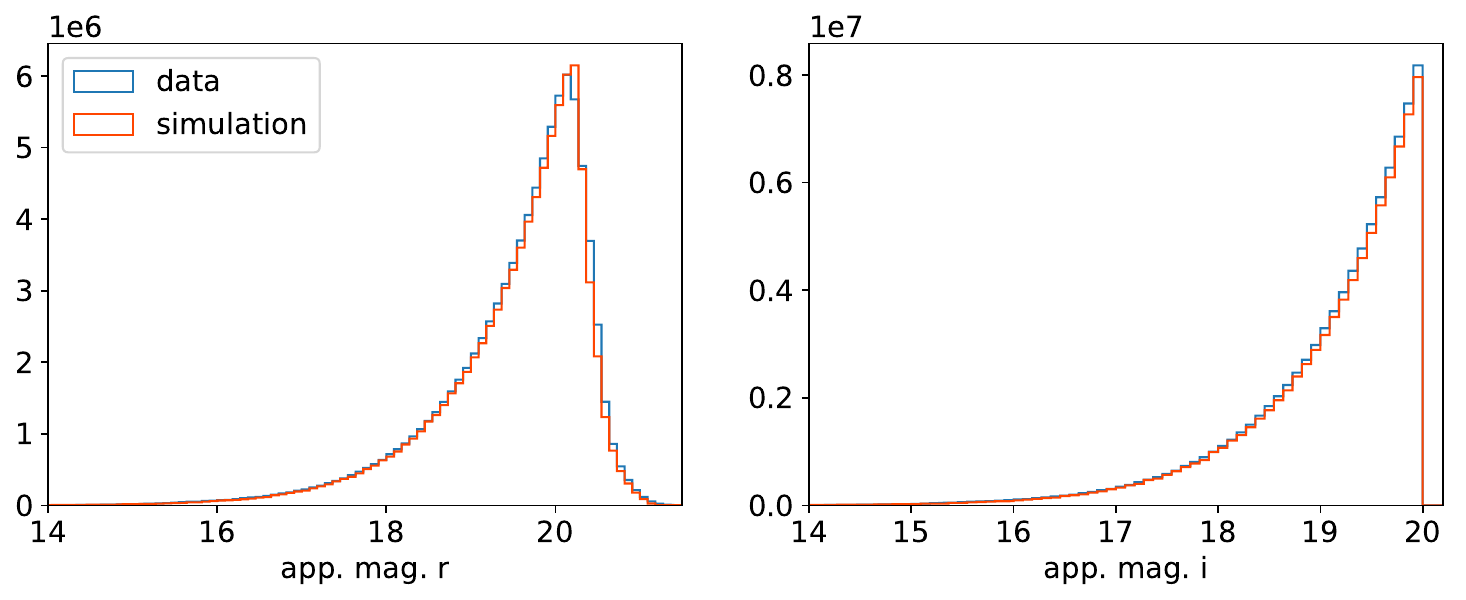}
\caption{Apparent magnitude histograms in the r and i band, for SHAM simulations (in orange) and the data (in blue).}
\label{fig:app_mag}
\end{figure}
\begin{figure}[h!]
\centering 
\includegraphics[width=.85\textwidth,angle=0]{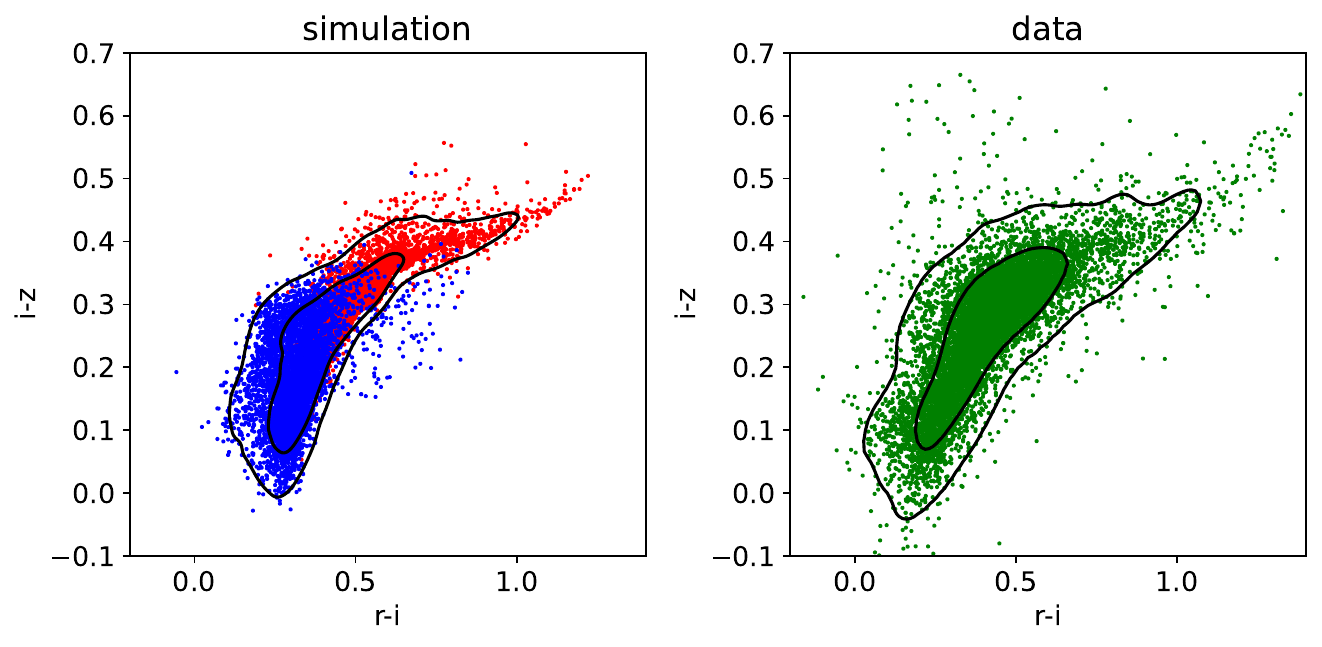}
\caption{Colour-colour scatter plot, comparing SHAM simulations (on the left) and the data (on the right). For the simulations, we additionally colour code the blue and the red galaxies, according to the two \texttt{UCat} galaxy populations. The lines indicate the 68\% and 95\% contours.}
\label{fig:color_space}
\end{figure}\\
\noindent Figure \ref{fig:color_space} displays the distribution of galaxies in colour space, for our simulation on the left and for the data on the right. Only a small number of galaxies is shown, for easier inspection. For the simulation, the catalogue is split into red and blue galaxies, according to the intrinsic types from \texttt{UCat}. The distribution in colour space agrees well with the data, though the data is more spread out and scattered due to uncertainties and scatter from \texttt{Source Extractor}. A possible way to improve this in the future could be to add scatter when applying the calibrated magnitude shift due to \texttt{Source Extractor}, but the scatter would have to be correctly correlated between the different magnitude bands.

\subsection{Galaxy clustering}
\label{sec:res_clustering}

\subsubsection{Spatial distribution}
\label{sec:spatial_distribution}

\begin{figure}[]
\centering 
\includegraphics[width=1.0\textwidth,angle=0]{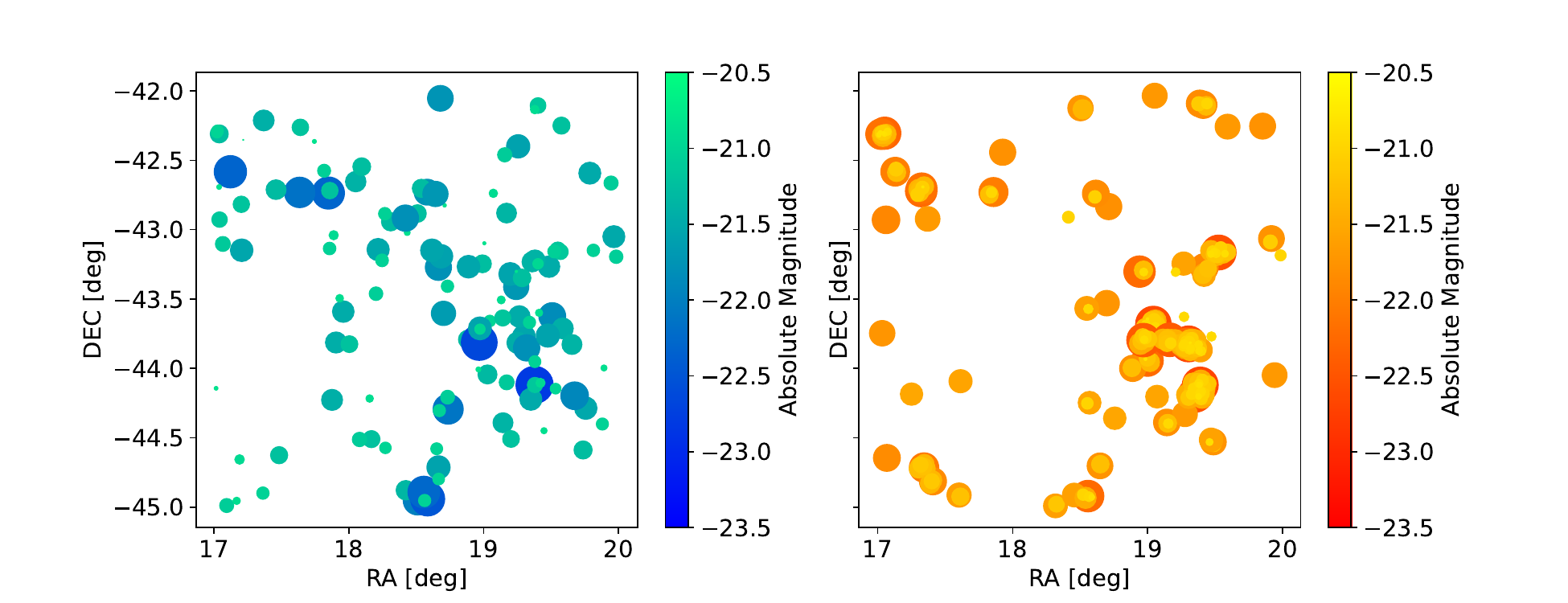}
\caption{Visualization of the clustering of blue (left) and red galaxies (right), within a sky area of about 6.5 $\deg^2$ in a narrow redshift range $0.40 < z < 0.41$. Galaxies are represented by circles, with both size and colour representing their absolute magnitudes.}
\label{fig:visualization_of_clustering}
\end{figure}

We assign the simulated galaxies differently to (sub-)halos depending on whether they are from the red or from the blue population within \texttt{UCat}. Therefore, the two populations do not have the same clustering. This can be seen in figure \ref{fig:visualization_of_clustering} where blue galaxies are shown on the left and red galaxies on the right. In both cases, we show the same cutout of the sky in right ascension and declination spanning about 6.5 $\deg^2$, and the same redshift range $0.40 < z < 0.41$. The simulated red galaxies are visibly more clustered, while the blue galaxies are more spread out to under-dense regions. This comes from the fact that we place red galaxies in highly clustered massive halos and into the majority of subhalos.
\begin{figure}[]
\centering 
\includegraphics[width=1.0\textwidth,angle=0]{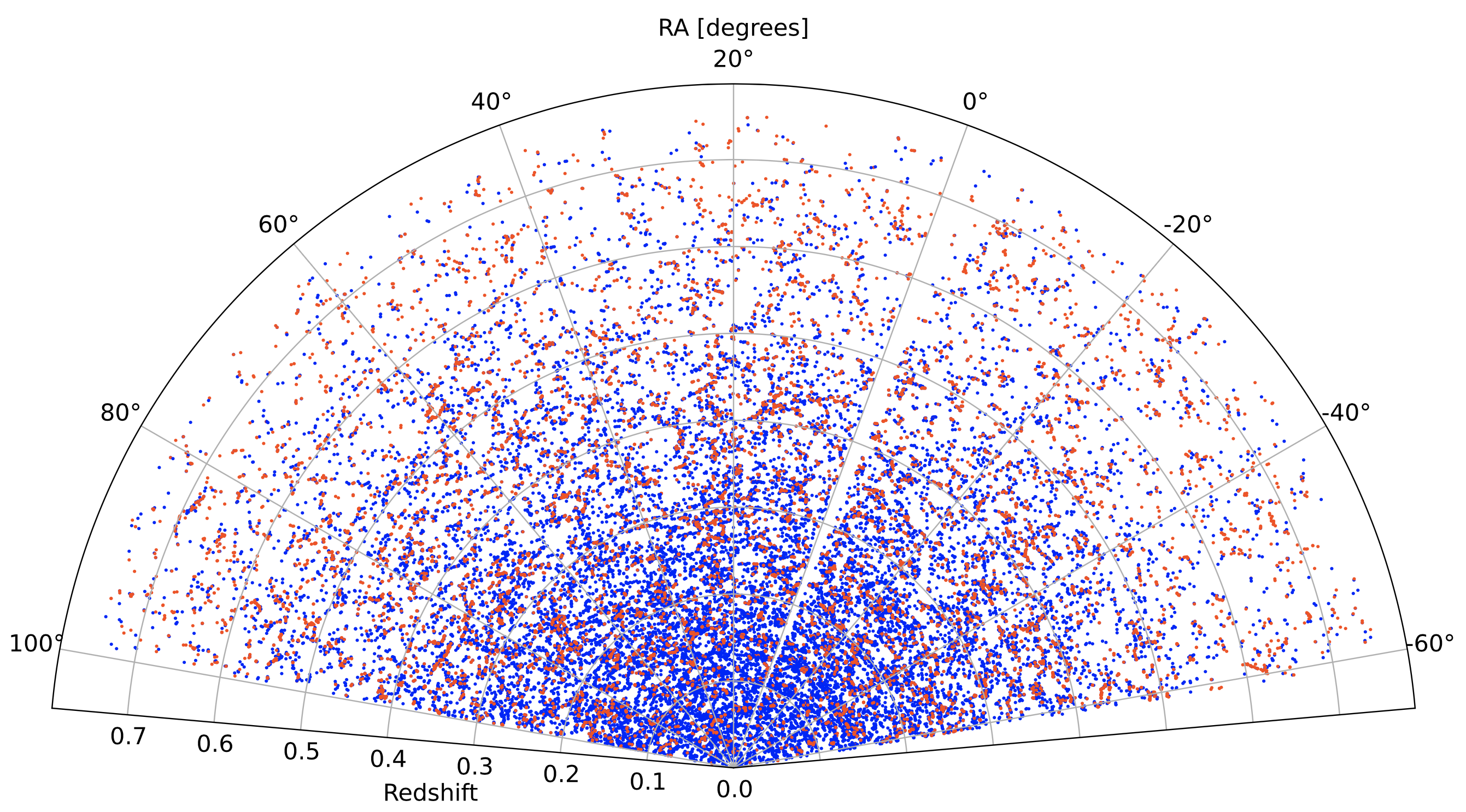}
\caption{Polar plot displaying the spatial galaxy distribution in real space as a function of redshift and angle, for both red and blue galaxies. Redshift is shown as the radial coordinate, and right ascension as the angular coordinate. For visual clarity, we only display galaxies within the declination range -55.2 deg < DEC < -55 deg.}
\label{fig:pie_plot}
\end{figure}\\
\noindent Figure \ref{fig:pie_plot} shows a visual representation of the resulting simulated galaxy clustering in the lightcone in real space. A cut-out through the lightcone along the redshift direction is shown, with redshift or comoving distance plotted as the radial coordinate. Only a small cut-out along DEC is shown for better visibility, -55.2 deg < DEC < -55. The structure on large scales including filaments, underdense and overdense regions is visible. Red and blue galaxies are shown in different colours.

\subsubsection{Correlation functions}
\label{sec:corr_func}

To compare the simulations with the data and to check whether they are in statistical agreement, we look at the two-point correlation function. We measure the angular correlation function $w(\theta)$ as a function of angular separation using \texttt{Corrfunc} \cite{Corrfunc_2019, Corrfunc_2020}. Given that we do not have radial information for the data, since DES in an imaging survey and therefore does not measure accurate redshifts, we cannot compare the 3D two-point correlation function. We therefore only consider the positions on the projected sphere.\\
We first evaluate the scales of trust in real space, in order not to over-interpret our simulations or the data on certain scales. We have defined certain physical scales corresponding to different effects in the dark matter simulations. This is related to the work presented in \cite{2021_berner}, where the simulations of halos and subhalos used here are described.\\
The following scales are in decreasing order. At 20 h$^{-1}$Mpc, we have medium scales where we trust our halo/subhalo simulations and they are well in agreement with the reference simulations based on \texttt{GADGET-2} with \texttt{Rockstar}. Below about 6 h$^{-1}$Mpc, \texttt{PINOCCHIO} starts to underestimate the clustering of halos due to its approximations. This scale is relatively inaccurate as it depends somewhat on the mass of the considered halos. The one-halo regime (according to the halo model of the power spectrum) starts at about 2 h$^{-1}$Mpc. While our simulations from \cite{2021_berner} added the subhalos to the output halo catalogues from \texttt{PINOCCHIO}, the density profile used to distribute the subhalos within halos could not be calibrated precisely. This is mainly because the subhalos from the numerical reference simulations had significant resolution effects. The work in \cite{2021_berner} saw a bump compared to the reference simulation, meaning a higher correlation function, at around 0.7 h$^{-1}$Mpc. Scales below 0.3 h$^{-1}$Mpc are too small to be simulated precisely with the given resolutions.
\begin{figure}[]
\begin{subfigure}{.48\textwidth}
  \centering
  % include first image
  \includegraphics[width=1.\linewidth]{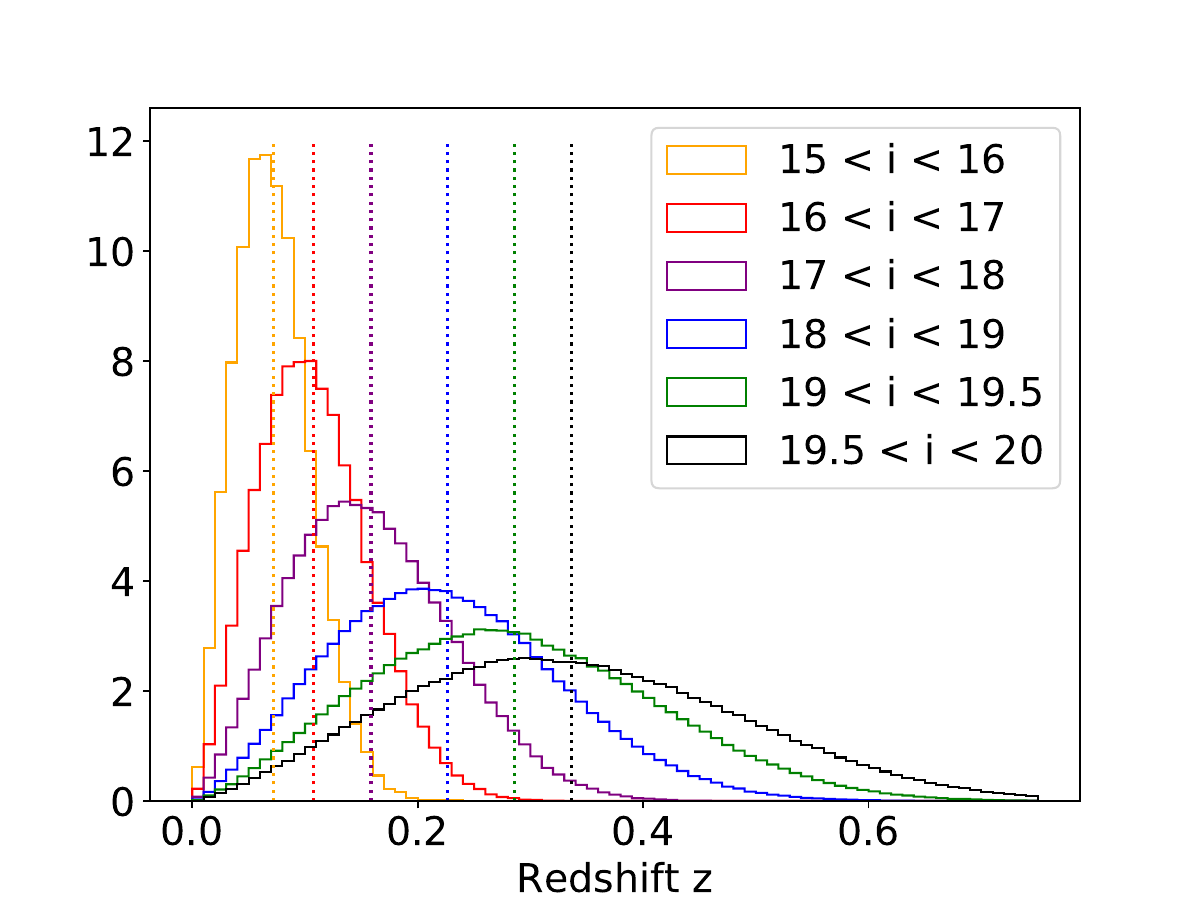}  
  \caption{n(z) for different magnitude bins}
  \label{fig:n_z_mag}
\end{subfigure}
\begin{subfigure}{.52\textwidth}
  \centering
  % include second image
  \includegraphics[width=1.\linewidth]{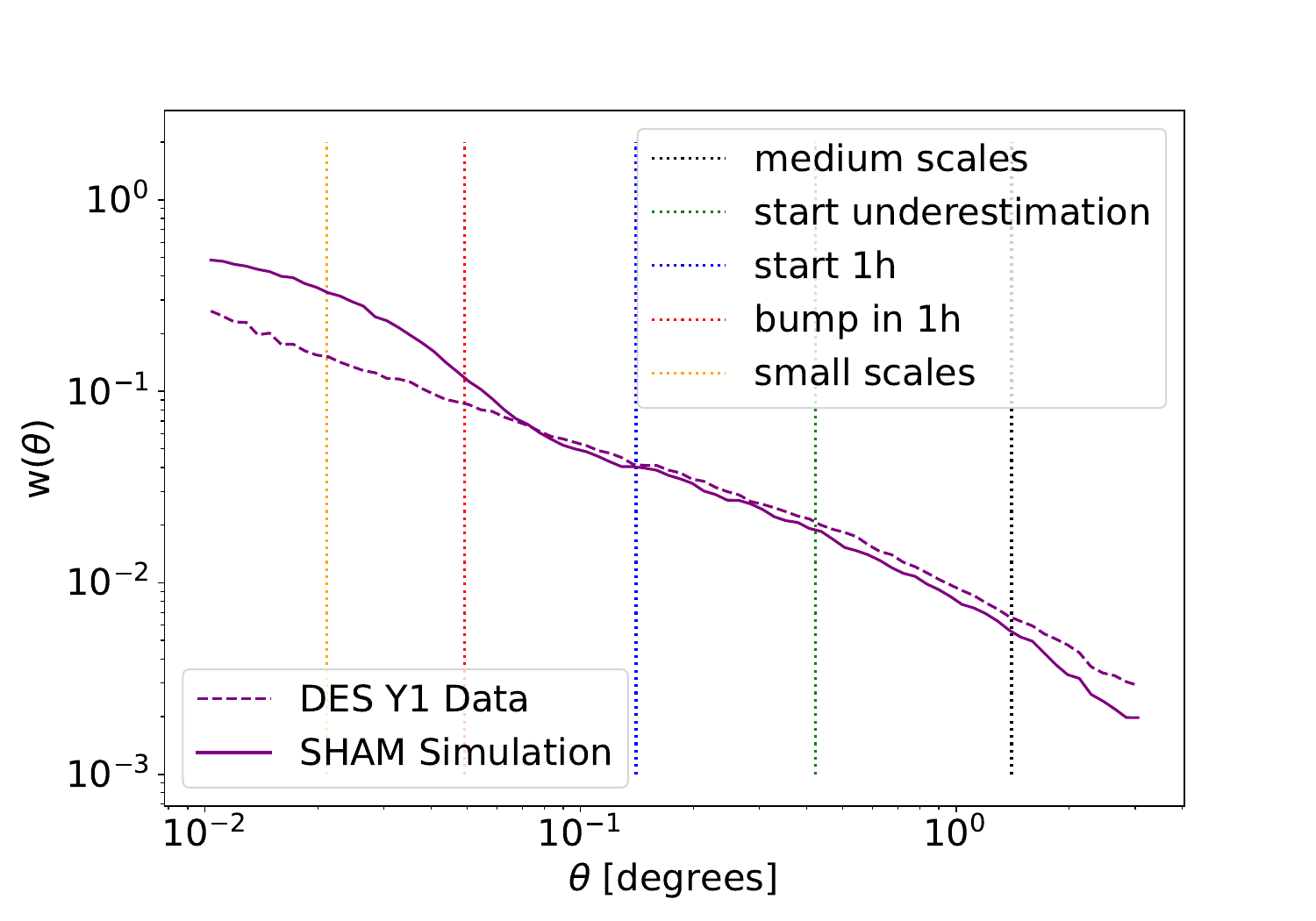}  
  \caption{Spatial scales of interest for $17 < i < 18$}
  \label{fig:2pcf_scales}
\end{subfigure}
\caption{Redshift distribution for different apparent magnitude bins on the left, calculated from the simulations. The dotted vertical lines are the mean redshifts for each magnitude bin. On the right, we show the two-point correlation function for data (dashed line) vs. simulation (solid line) for the magnitude bin $17 < i < 18$, to indicate certain scales of interest (dotted vertical lines).}
\label{fig:scales_of_trust}
\end{figure}\\
\noindent To relate these physical scales to our results, we transform them from comoving distances to angular distances projected on the sky. As this relation depends on redshift, figure \ref{fig:n_z_mag} shows the redshift distributions of the simulated galaxies for different magnitude bins. Given that we place the galaxies drawn from \texttt{UCat} into halos and subhalos, each resulting galaxy has a precise redshift in the simulated lightcone. The dotted vertical lines in figure \ref{fig:n_z_mag} indicate the mean redshifts for each magnitude bin.\\
For the magnitude bin $17 < i < 18$, we show the resulting scales in figure \ref{fig:2pcf_scales}, along with the two-point correlation functions from the data and from one simulation as a reference. Given the discussion above, we can only use scales within the two-halo regime for a stable comparison between data and simulations. For the magnitude bin shown in \ref{fig:2pcf_scales}, this corresponds to angular scales with $\theta > 0.15$ degrees. Given that brighter magnitude bins have a lower mean redshift than fainter magnitude bins, the usable scales correspond to smaller angles for fainter galaxies. This effect would only be increased if the mass dependence of the physical scales were included, since brighter galaxies sit in larger halos. For the following comparison, we only show the scales within the two-halo regime for each magnitude bin.
\begin{figure}[h!]
\centering 
\includegraphics[width=0.999\textwidth,angle=0]{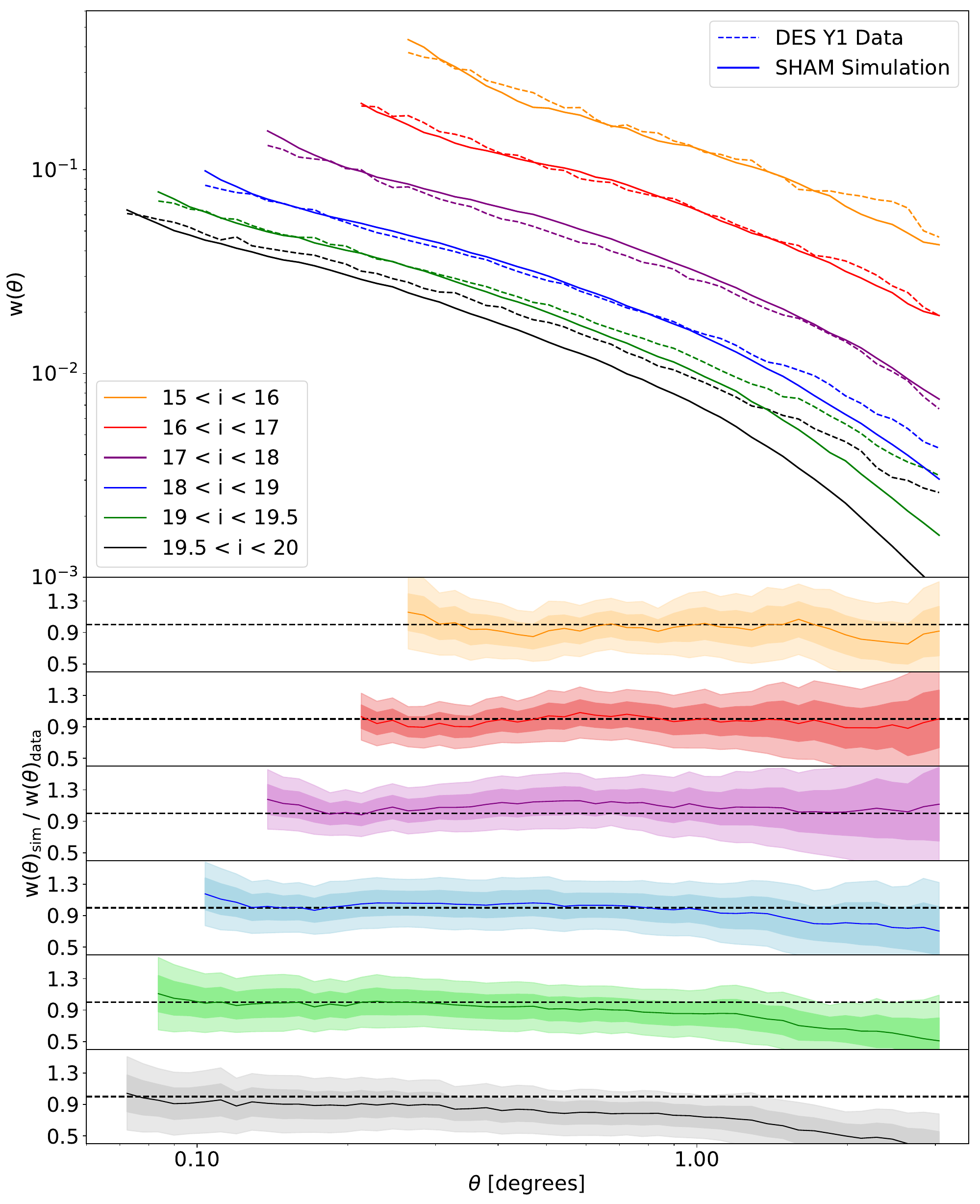}
\caption{Galaxy two-point correlation function, as a function of separation angle $\theta$, projected along the line of sight. Solid lines are the mean values from 10 simulations, dashed values correspond to the data. Different colours refer to different bins in apparent magnitude i. The lower panels show the ratio of simulations to data, with the 1$\sigma$ and 2$\sigma$ intervals due to statistical and systematic uncertainties.}
\label{fig:2pcf}
\end{figure}\\
In figure \ref{fig:2pcf}, we show the comparison of galaxy clustering between data and simulations. The data is shown as dashed lines, while the simulations are shown as solid lines. Different colours correspond to different magnitude bins in the i band. For the simulations, the mean from 10 \texttt{PINOCCHIO} realizations with the same SHAM model is shown. For this figure, the fiducial SHAM parameters were chosen as M$_{\mathrm{limit}} = 8 \cdot 10^{12}$ h$^{-1}$M$_{\odot}$ and t$_{\mathrm{quench}} = $ 2.0 Gyr, since t$_{\mathrm{quench}}$ is expected be about 2 Gyr for redshifts $0<z<1$ and the mass limit between $10^{12}$ and $10^{13}$ h$^{-1}$M$_{\odot}$ from literature \cite{2010_book_mo}.\\
In the lower panels, the ratio between simulations and data in the corresponding bands is shown. The coloured areas given are the 1$\sigma$ and 2$\sigma$ intervals from our estimation of uncertainties. We combined the statistical and systematic uncertainties together. Statistical variability comes from running different \texttt{PINOCCHIO} runs. Systematic uncertainty comes here from the galaxy population model, where we run the same simulation with a fixed set of SHAM parameters and different posterior points from \cite{Herbel_2017}. The parameters include shape and normalization of the luminosity function, which directly affect the mass to luminosity relation derived with our SHAM method.\\
We find that our simulations are well in agreement with the data, mostly within 1$\sigma$. The least agreement is achieved for the highest magnitude bin, meaning the faintest galaxies, for which the simulation slightly under-predicts the clustering, especially for larger separation angles.
\begin{figure}[]
\begin{subfigure}{.5\textwidth}
  \centering
  % include first image
  \includegraphics[width=1.\linewidth]{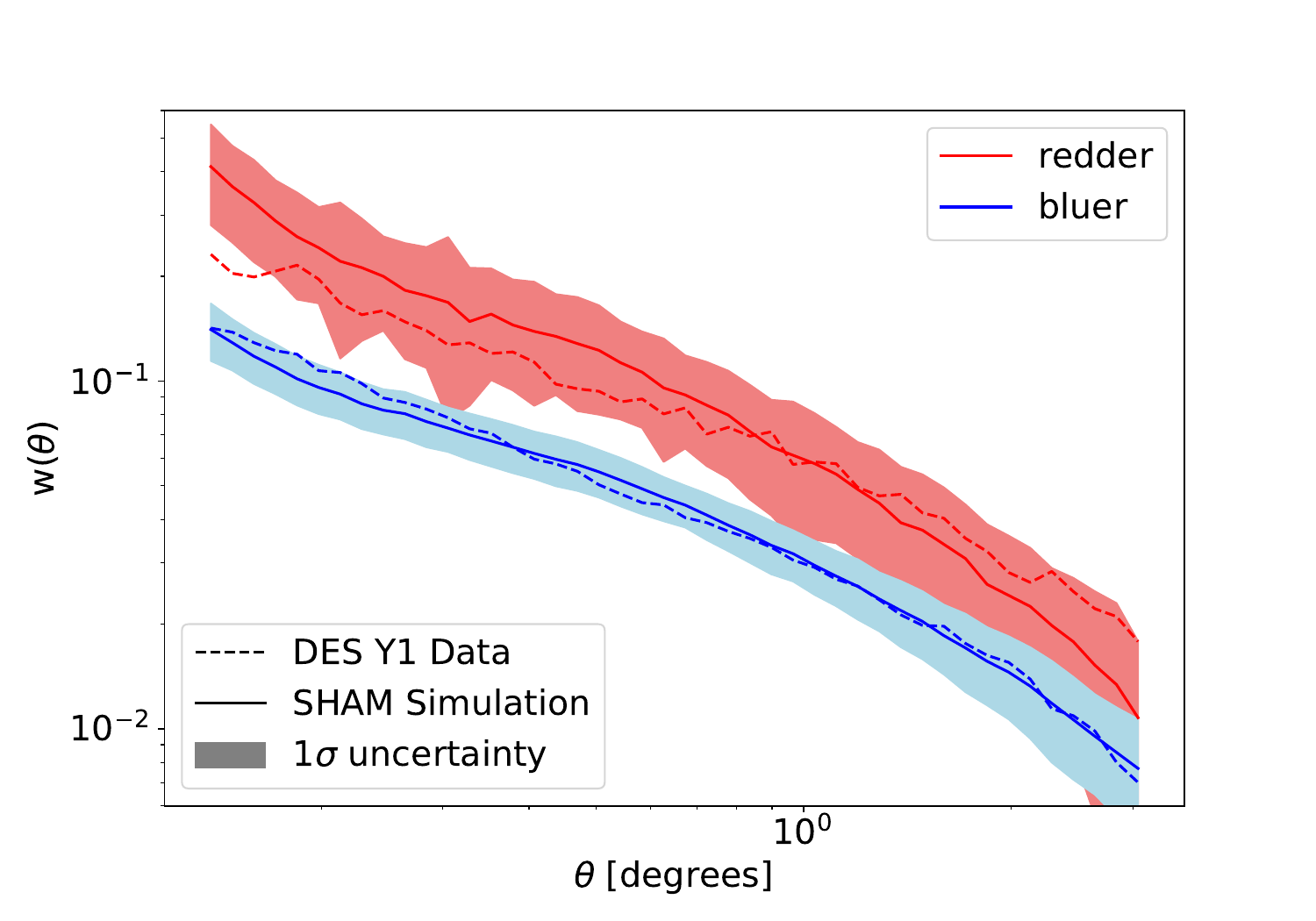}  
  \caption{$17 < i < 18$}
  \label{fig:2pcf_redblue_1718}
\end{subfigure}
\begin{subfigure}{.5\textwidth}
  \centering
  % include second image
  \includegraphics[width=1.\linewidth]{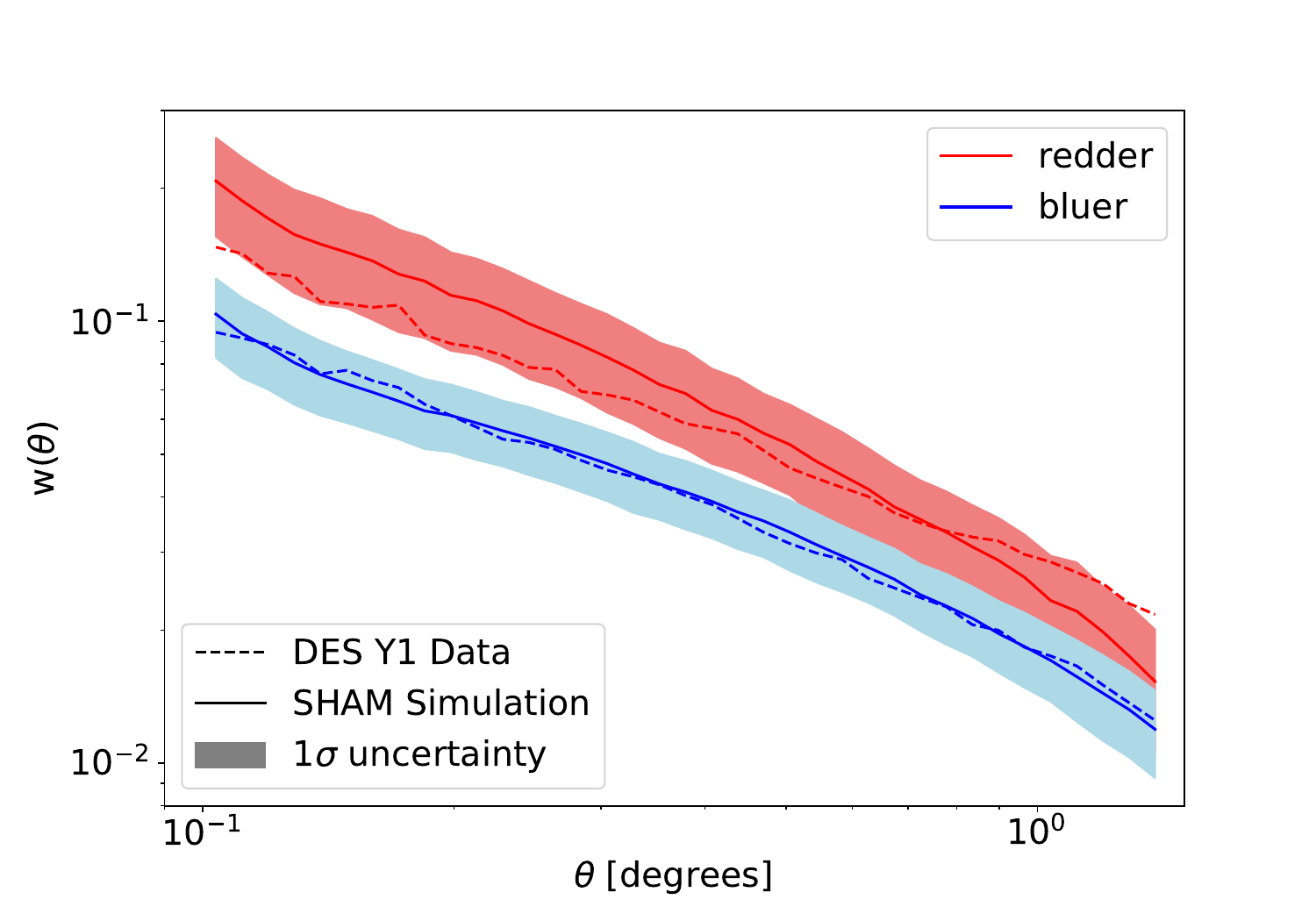}  
  \caption{$18 < i < 19$}
  \label{fig:2pcf_redblue_1819}
\end{subfigure}
\caption{Angular two-point correlation functions for redder ($i-z > 0.3$) and bluer galaxies ($i-z < 0.3$), comparing our simulations to the data. For the simulations, the mean values from 10 simulation runs are shown. The 1$\sigma$ ranges correspond to the combination of the statistical and systematic uncertainties in the simulations. On the left, the magnitude bin $17 < i < 18$ is shown, while on the right, we present the bin $18 < i < 19$.}
\label{fig:2pcf_redblue}
\end{figure}\\
\noindent Our SHAM model with its two free model parameters distinguishes between red and blue galaxies from \texttt{UFig}. The red vs. blue label of the simulated galaxies is not an observable property, but we can define a cut in colour space to perform a colour dependent comparison between our simulations and the DES data.  A colour cut allows us to consistently separate the simulations and the data into a "redder" and a "bluer" subset, making it possible to measure clustering separately. The cut we use for this separation is a straight line at $i-z = 0.3$, with redder galaxies having $i-z > 0.3$ and bluer galaxies having $i-z < 0.3$. The reason for this choice is that it approximately splits the simulation into the red and blue galaxies according to the categories from \texttt{UFig}. Furthermore, we chose a simple cut involving few magnitude bands, to minimize the effect from differences in colour space (see figure \ref{fig:color_space}).\\
As described in section \ref{sec:mag_cal}, we have to carefully calibrate any shift in the magnitudes due to \texttt{Source Extractor}. For the analysis with redder and bluer galaxies, we have first calibrated the shift in $i-z$, before cutting into the two samples. Afterwards, starting with the original $i$-band magnitudes, we have calibrated the shift in $i$ separately for redder and bluer galaxies. This allows us to bin the two catalogues into magnitude bins and compare the correlation functions from the data with our simulations, similar to figure \ref{fig:2pcf}. The reason for performing a separate calibration for redder and bluer galaxies is that while the magnitude dependence of the shift in each magnitude band is small, the resulting colour dependence of the magnitude shift is significant.\\
In figure \ref{fig:2pcf_redblue}, we show the angular two-point correlation functions for redder and bluer galaxies. We show the results for two separate magnitude bins, $17<i<18$ (subfigure \ref{fig:2pcf_redblue_1718}) and $18<i<19$ (subfigure \ref{fig:2pcf_redblue_1819}). The correlation functions are calculated using \texttt{Corrfunc}. The dashed lines represent the result from the DES data, while the solid lines are the mean values from 10 simulation runs with the fiducial SHAM parameters. The shaded areas show the 1$\sigma$ uncertainties, estimated by combining the statistical uncertainties with the variations from different posterior points, as described above for figure \ref{fig:2pcf}.\\
The simulations agree with the data both for the redder and the bluer subset. The colour dependent clustering agrees better for the bluer galaxies, since most redder galaxies occupy subhalos, which are simulated with less stability than the halos at a given resolution. We do not show the brighter magnitude bins, as there are few redder galaxies below $i=17$. The brighter bins are in agreement due to the large scatter between different simulation runs and especially between different posterior points, but do not provide much insight.\\
Such a colour-dependent analysis is limited by the accuracy of the underlying galaxy model. As shown in figure \ref{fig:color_space}, the posterior from the ABC analysis in \cite{Herbel_2017} gives a good but not perfect match to the data in colour space. A cut in colour space therefore results in a slightly different sample for the simulations and the data, more prominently than a simple magnitude cut. Future works may be able to distinguish between redder and bluer galaxies more accurately, by using newer ABC posteriors resulting in a better agreement with the data. It is worth mentioning that we consider a bright sample of galaxies, which contributes only a small fraction of the whole galaxy population used for the ABC analysis.

\subsection{Galaxy to halo relation}
\label{sec:gal_halo_rel}

Our main goal is to build the SHAM model and show that the resulting clustering agrees with observational data. Since the relation between galaxy luminosities and halo masses is derived when the halos and subhalos are matched to red and blue galaxies, it is not assumed or an input. In this section we present the effect of the SHAM parameters, as well as the resulting galaxy-halo relation when using the reference SHAM parameters.

\subsubsection{Constraints on the SHAM parameters}
\label{sec:constraints}
\begin{figure}[t]
\centering 
\includegraphics[width=0.8\textwidth,angle=0]{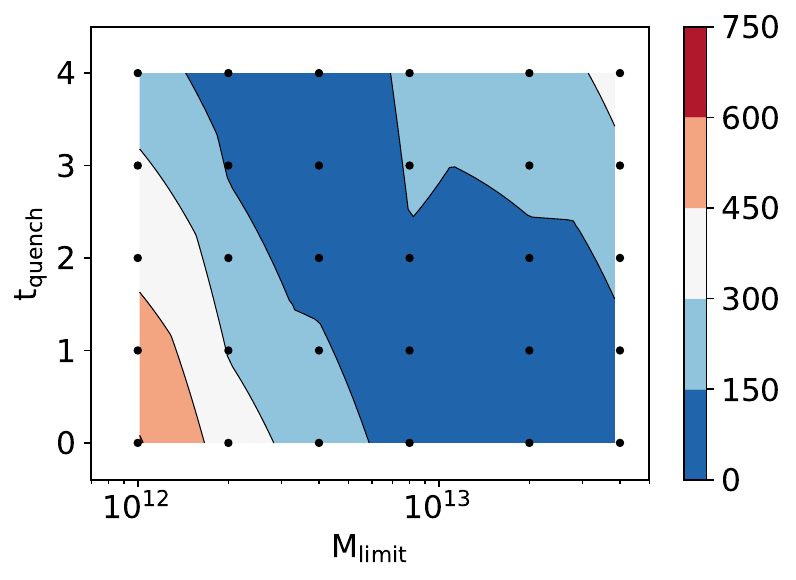}
\caption{The resulting chi-squared surface for the two SHAM model parameters M$_{\mathrm{limit}}$ on the x-axis and t$_{\mathrm{quench}}$ on the y-axis. The black dots show the grid points at which we run a simulation. There is a degeneracy from the top left to the bottom right.}
\label{fig:chi2_surface}
\end{figure}

We have calculated the covariance matrix for the simulations based on 10 simulation runs with the same cosmological and SHAM parameters but different underlying \texttt{PINOCCHIO} runs. For this and the following chi-squared analysis, we have calculated the angular two-point correlation functions $w(\theta)$ using \texttt{Corrfunc}. We split up the data set into 5 magnitude bins: $15<i\leq16$, $16<i\leq17$, $17<i\leq18$, $18<i\leq19$ and $19<i\leq20$. For better stability, we have reduced the number of bins in angular separation $\theta$. Since the scales of trust depend on apparent magnitude, as discussed in section \ref{sec:corr_func}, this gave us 4 to 6 bins in $\theta$ between about $0.1\deg$ and $30\deg$, with the number of bins depending on the magnitude bin.\\
Using the same bins as for the simulations, both in magnitude and angular separation, we calculated the correlation functions for the observational data. We then computed the chi-squared values using $\chi^2 = D^{T}COV^{-1}D$, where $D$ is the data vector defined as $D = w_{\mathrm{data}} - w_{\mathrm{sim}}$. Both for the data vector and for the covariance matrix $COV$ the $w(\theta)$ vectors for the separate magnitude bins are stacked together. Since the covariance matrix is ill-conditioned, we used Tikhonov Regularization \cite{tikhonov1977solutions, tikhonov1995numerical} for the inversion. Figure \ref{fig:chi2_surface} shows the grid points of the SHAM parameters we used, as well as the contours of the chi-squared surface. The chi-squared values are not normalized here.
\\
\noindent We investigated values between 0 and 4 Gyr for t$_{\mathrm{quench}}$, as we expect it to be around 1-2 Gyr for redshifts $0<z<1$ (e.g. \cite{2010_book_mo}). Values of t$_{\mathrm{quench}}$ below 0 Gyr are impossible, and values above 4 Gyr would make almost all satellite galaxies in clusters red, which is not what observations show.\\
For M$_{\mathrm{limit}}$, we considered the range $10^{12}$ to $4 \cdot 10^{13}$ h$^{-1}$M$_{\odot}$. Literature places the transition between red and blue centrals in the range $10^{12}$ to $10^{13}$ h$^{-1}$M$_{\odot}$ \cite{2010_book_mo}. Values of M$_{\mathrm{limit}}$ above $2 \cdot 10^{13}$ h$^{-1}$M$_{\odot}$ lead to too few red central galaxies, inconsistent with observations. On the other hand, values below $10^{12}$ h$^{-1}$M$_{\odot}$ would mean that most medium sized galaxies should be red, again disagreeing with observations.\\
As discussed in more details in section \ref{sec:impact_params}, increasing either the quenching time t$_{\mathrm{quench}}$ for satellite galaxies or the mass limit M$_{\mathrm{limit}}$ for central galaxies has a similar effect on galaxy clustering. This is the reason for the degeneracy between the two parameters, as can be seen in figure \ref{fig:chi2_surface}. The best fit parameters lie along the degeneracy between (M$_{\mathrm{limit}}$, t$_{\mathrm{quench}}$) = ($2 \cdot 10^{12}$, 4.0), and ($4 \cdot 10^{13}$, 0.0). The units are h$^{-1}$M$_{\odot}$ for the mass and Gyr for the time. The constraining power is relatively weak since the parameters do not have a strong effect on clustering. Therefore, the shape of the best fit contour is irregular, and its size is large. The exact shape also depends on the number of layers used for plotting, but the degeneracy remains the same.

\subsubsection{Impact of the SHAM parameters on clustering}
\label{sec:impact_params}
The faintest galaxies sit in the smallest halos, by definition of abundance matching, which have the lowest correlation function. This holds mainly for halos and not for subhalos, since subhalos are much more clustered. On small scales, subhalos and therefore satellite galaxies are more clustered due to the 1-halo term. On large scales, subhalos are more clustered compared to halos of the same mass since they sit in host halos which are often 100 times as massive as themselves, and more massive halos are more clustered.\\
To understand the effect of the two SHAM parameters, it is important to note that the galaxy sample we look at is cut in magnitude. The whole simulation is limited in the i-band by $i<20$, and each magnitude bin corresponds to two cuts according to its bin limits. Therefore, although each halo and subhalo is populated with a galaxy, many of them are not included. Galaxies are excluded if they are too faint for the magnitude cut. For a given redshift, galaxies are effectively cut in absolute magnitude or luminosity. Overall, distant galaxies appear much fainter and are therefore excluded unless they are very bright.\\
Increasing the mass limit M$_{\mathrm{limit}}$ leads to blue galaxies occupying larger central halos. Since the luminosity function of blue galaxies is fixed, this leads to small halos being occupied by fainter galaxies, thus leading to them being excluded by the selection. Since small halos lower the correlation function, this increases the overall clustering.\\
A similar effect happens when the quenching time t$_{\mathrm{quench}}$ is increased. This leads to more blue galaxies occupying subhalos and becoming satellites. Therefore, the small halos are filled with faint galaxies, making them be removed by the magnitude cuts. In either case, to increase clustering one needs to ensure that small halos are not included, therefore increasing M$_{\mathrm{limit}}$ and/or t$_{\mathrm{quench}}$.

\begin{figure}[]
\begin{subfigure}{.49\textwidth}
  \centering
  % include second image
  \includegraphics[width=1.\linewidth]{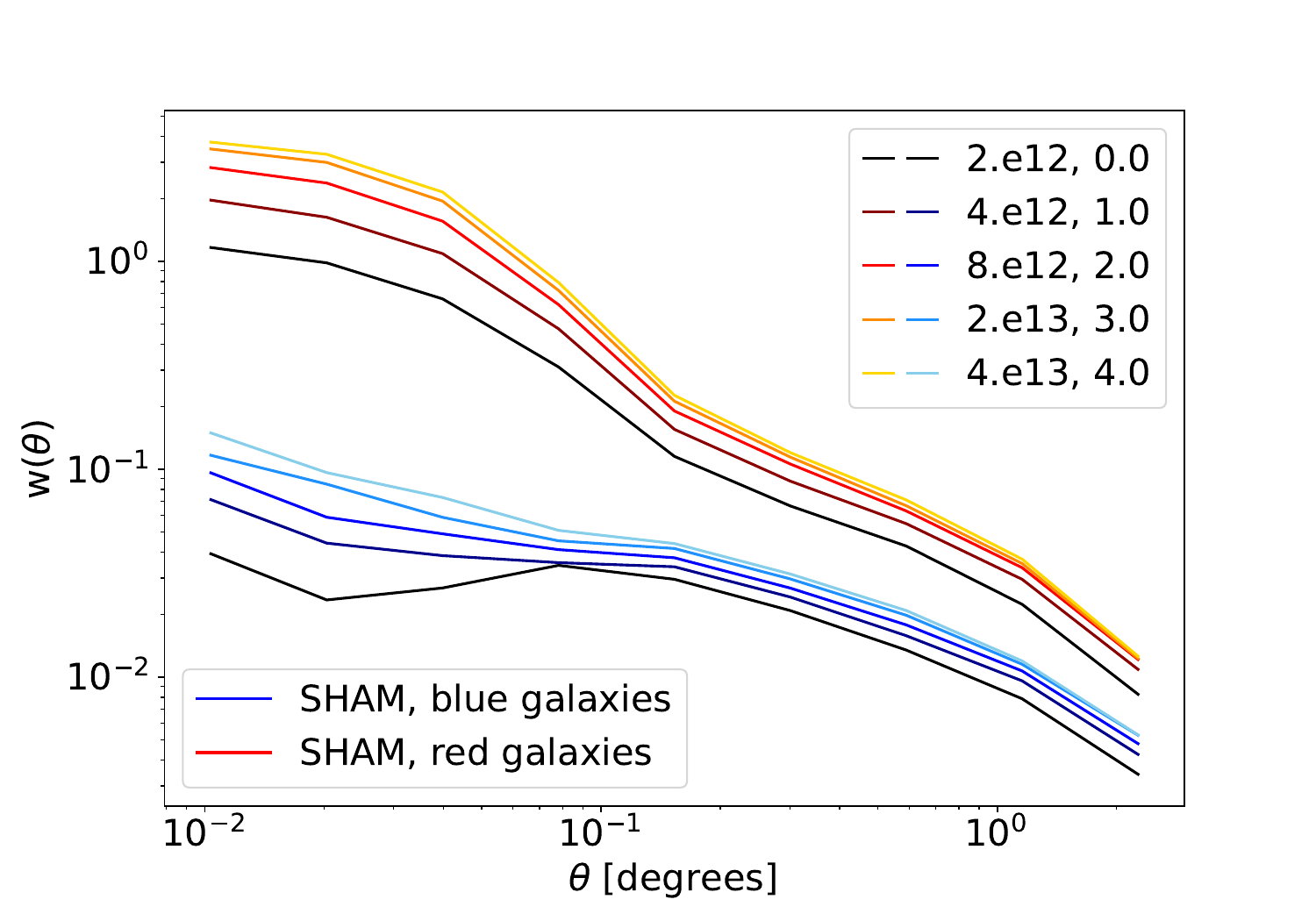} 
  \caption{Perpendicular to the degeneracy}
  \label{fig:2pcf_perp_deg}
\end{subfigure}
\begin{subfigure}{.49\textwidth}
  \centering
  % include first image
  \includegraphics[width=1.\linewidth]{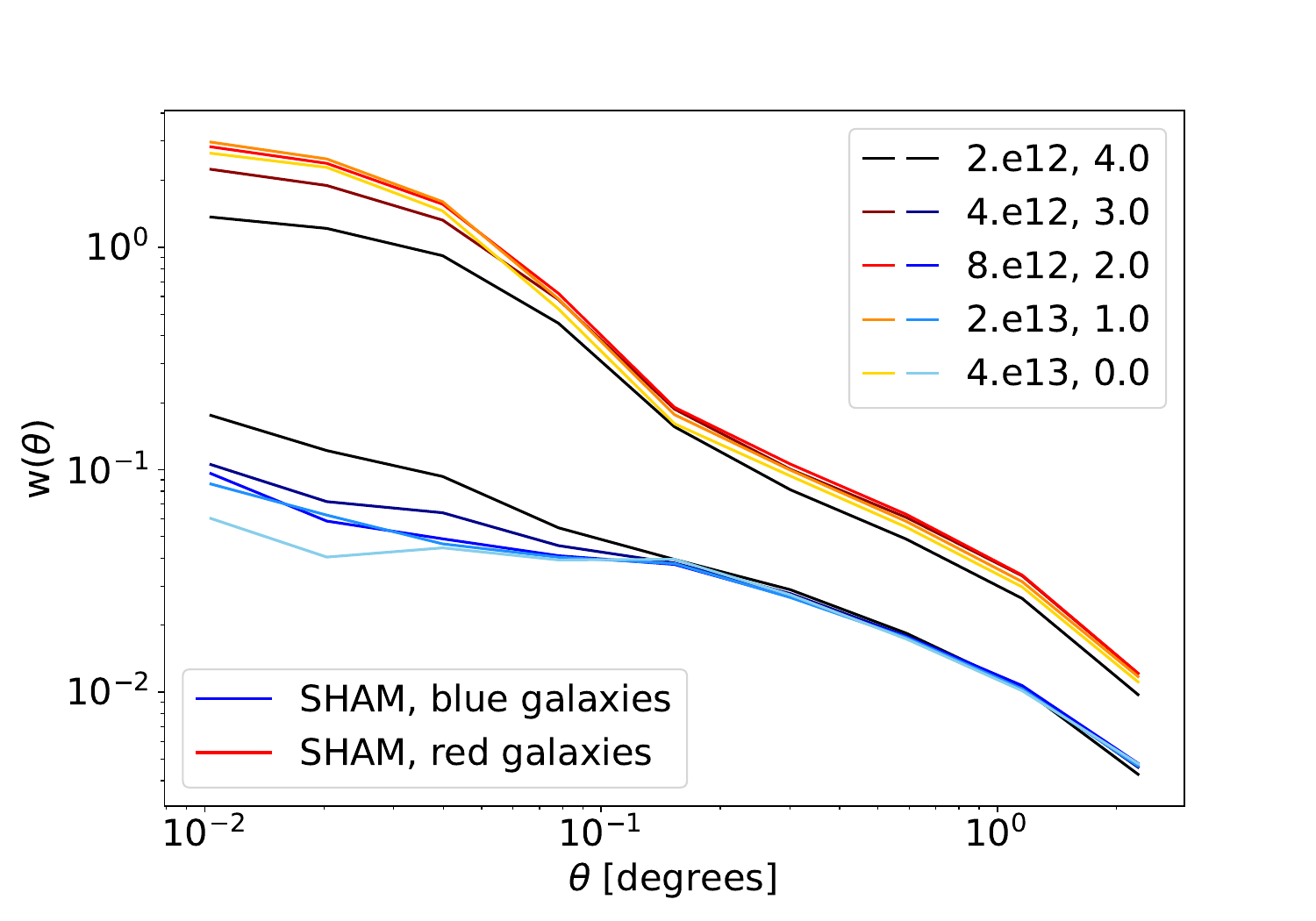}  
  \caption{Along the degeneracy}
  \label{fig:2pcf_along_deg}
\end{subfigure}
\caption{Both figures show the clustering of red vs. blue galaxies for apparent magnitudes within $18 < i < 19$. The different shades of red and blue correspond to different points in the parameter space, with M$_{\mathrm{limit}}$ in h$^{-1}$M$_{\odot}$ being given first in the legend, then t$_{\mathrm{quench}}$ in Gyr. On the left, the dependence of galaxy clustering across the degeneracy shows an increase in clustering on all scales with increasing M$_{\mathrm{limit}}$ and t$_{\mathrm{quench}}$. On the right, the dependence along the degeneracy is shown, where the effect is mostly on small scales.}
\label{fig:2pcf_effect_params}
\end{figure}

\noindent This can be seen in figure \ref{fig:2pcf_perp_deg}, where the correlation function for the magnitude bin $18<i<19$ is shown. Only results from our simulations are shown here, and we distinguish between red and blue galaxies according to the two \texttt{UCat} populations. The values in the 2D parameter plane for this figure were chosen perpendicular to the degeneracy. The colours from dark to light shades are from small to high values of M$_{\mathrm{limit}}$ and t$_{\mathrm{quench}}$. The clustering varies on all scales for both red and blue galaxies, with the effect being strongest in the 1-halo regime.\\
The relation in figure \ref{fig:2pcf_along_deg} is analogous to figure \ref{fig:2pcf_perp_deg}, but with the parameter points being chosen along the degeneracy. Since increasing M$_{\mathrm{limit}}$ and decreasing t$_{\mathrm{quench}}$ or vice versa cancels out the effect of the parameters on clustering, the 2-halo term in figure \ref{fig:2pcf_along_deg} is almost unchanged. On small scales, in the 1-halo regime, the overall correlation function would also be mostly unchanged. Since we split the sample into red and blue galaxies, we can see an effect. A low value of t$_{\mathrm{quench}}$ leads to few or no blue galaxies being satellites, therefore strongly reducing the 1-halo term. Many red satellites surviving the cuts on the other hand increases the clustering of red galaxies on small scales.

\subsubsection{Mass to luminosity relation}
\label{sec:mass_lum_rel}
\begin{figure}[]
\begin{subfigure}{.49\textwidth}
  \centering
  \includegraphics[width=1.\linewidth]{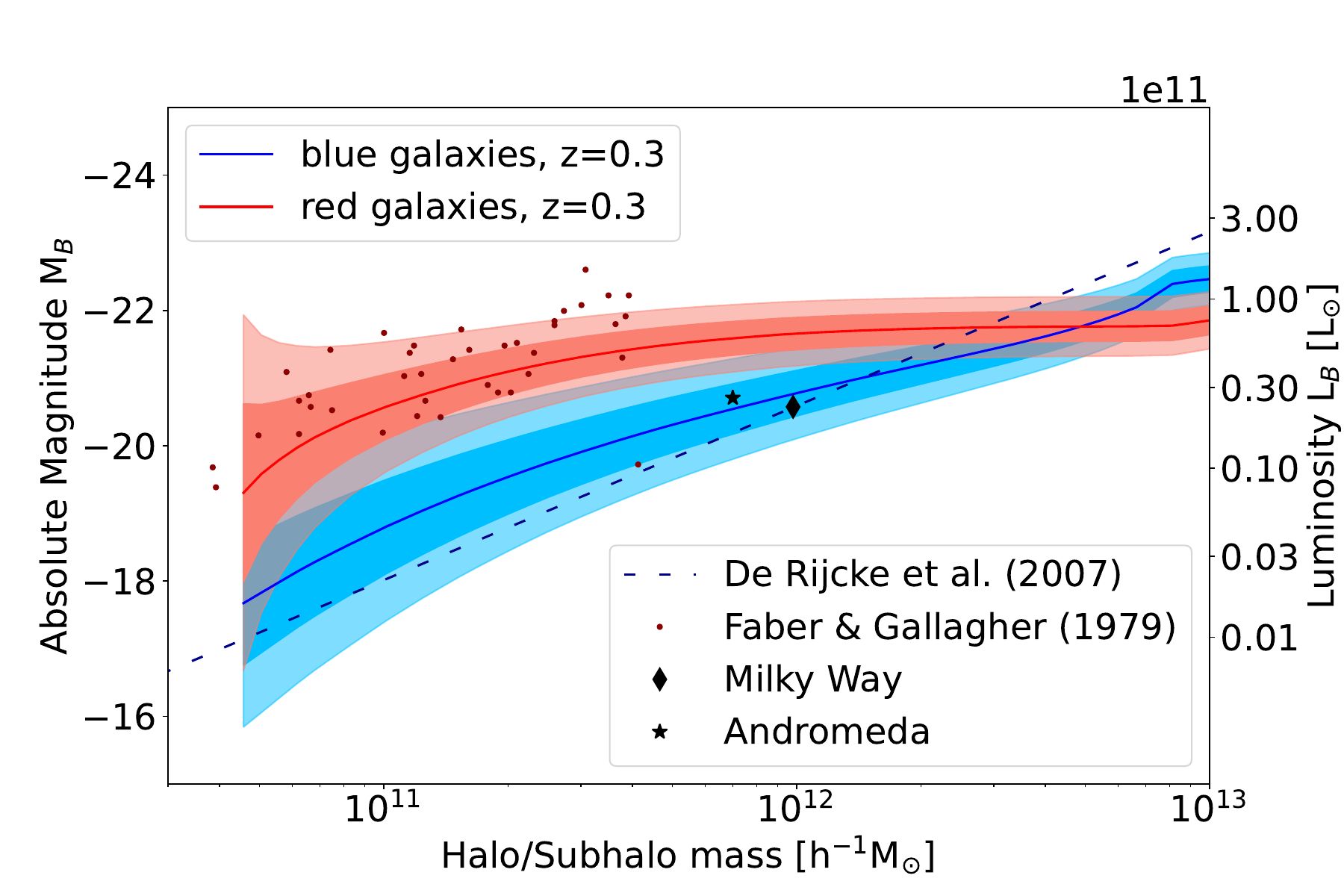}  
  \caption{Mass to luminosity relation}
  \label{fig:mass_abs_mag}
\end{subfigure}
\begin{subfigure}{.49\textwidth}
  \centering
  \includegraphics[width=1.\linewidth]{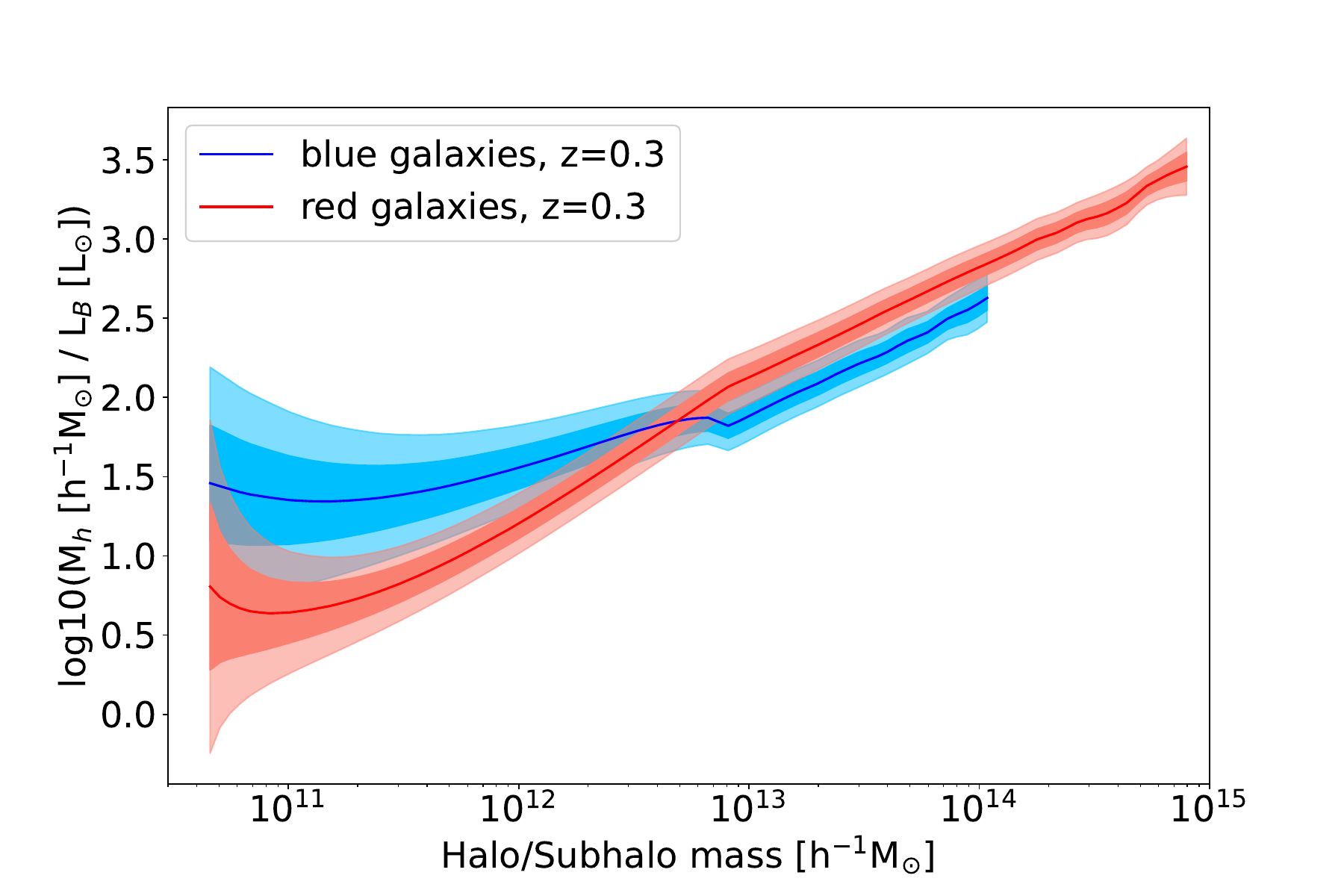}  
  \caption{Relation between mass and mass-to-light ratio}
  \label{fig:mass_masstolight}
\end{subfigure}
\caption{Relation between (sub)halo mass and luminosity, derived using a two-dimensional interpolation in the mass-redshift plane. Red and blue galaxies have separate relations. On the left, the relation between mass and absolute magnitude or luminosity is shown. The points are observed galaxies of different types, from other works \cite{1979_faber, 1999_bergh, 2018_kafle, 2019_watkins}. The dashed line is from a Tully-Fisher relation for observed spiral galaxies \cite{DeRijcke_2007}. On the right, the mass-to-light ratio is shown in relation to the (sub)halo mass. The relations are shown at redshift z=0.3. The shaded areas correspond to the 1$\sigma$ and 2$\sigma$ uncertainties.}
\label{fig:mass_lum}
\end{figure}

A natural way to show the relation between halos and galaxies for a SHAM model is to look at the relation between the two properties that are matched. In our case, the two properties are the halo/subhalo mass and the galaxy absolute magnitude or luminosity. We show the mass to luminosity relation for red and blue galaxies separately for our reference SHAM parameters in subfigure \ref{fig:mass_abs_mag}. The relation depends on redshift, and we show it for redshift z=0.3, which is close to the mean redshift for our simulations. The non-smooth feature in the relation comes from the fact that the mass limit M$_{\mathrm{limit}}$ leads to halos only contributing above the limit for red galaxies and only below the limit for blue galaxies. Subhalos are split according to t$_{\mathrm{quench}}$, but this is not a cut in mass and therefore not directly visible in the mass to luminosity relation. In subfigure \ref{fig:mass_masstolight}, we show the same relation, but with the mass-to-light ratio on the y-axis.
 \\
\noindent When comparing to other works, we find that our resulting relation between galaxies and halos agrees with literature. In subfigure \ref{fig:mass_abs_mag}, we show data for local elliptical galaxies from \cite{1979_faber} as dark red dots, which is in agreement with our result for red galaxies. We also show data points for the Milky Way and Andromeda (values from \cite{1979_faber, 1999_bergh, 2018_kafle, 2019_watkins}, in black), two star forming spiral galaxies, in agreement with our relation for blue galaxies. Spiral galaxies have more active star formation than elliptical ones. While the correspondence is not exact, we compare observational data for elliptical galaxies to our red galaxies and data for spiral galaxies to our blue ones. Note that our relation depends on redshift, and only a specific redshift of $z=0.3$ is shown here, while the observed galaxies are at various redshifts around this value. The redshift dependence of the mass to luminosity relation is significant overall but weak for a narrow redshift range.\\
We further show a linear fit between log circular velocity $\log(v_c)$ and log luminosity $\log(L_B)$ for the Tully-Fisher relation (see e.g. \cite{1977_tf}) for spiral galaxies from \cite{DeRijcke_2007} as a dashed line. We relate $v_c$ to (sub-)halo mass $M$ via $v_c = (10 ~ G H(z) M )^{1/3}$. This Tully-Fisher relation is in good agreement with our mass to luminosity relation for blue galaxies.\\
The data from \cite{2005_wel} on early-type galaxies at $z \sim 1$ is also in agreement with our mass to luminosity relation for red galaxies at $z=1$, but not shown for better visibility. Other studies (e.g. \cite{2010_behroozi, 2013_reddick}) show similar trends as our relation, but are not directly comparable, as they measured luminosities in other bands or stellar masses.

\section{Conclusion}
\label{sec:conc}
We presented our galaxy forward model based on Subhalo Abundance Matching (SHAM). The halo lightcones are generated with PINOCCHIO. The subhalo extraction from the merger history presented in \cite{2021_berner} was used to add the surviving subhalos to the halo catalogue. This allows us to create halo-subhalo lightcone catalogues at high resolution with large cosmological volumes.\\
The galaxies used to populate the halos and subhalos are drawn from luminosity functions from \texttt{UCat}, the catalogue component of the Ultra-Fast Image Generator (\texttt{UFig}). \texttt{UCat} contains two separate galaxy populations, for red and blue galaxies. The blue galaxies have more star formation, while the red galaxies are more quiescent.\\
These two galaxy populations are included in our SHAM model based on star formation quenching. With time, the formation of stars is quenched due to several processes (e.g., reduced inflow of gas or tidal stripping), making blue galaxies become red. For central galaxies, our SHAM model includes mass quenching, with the parameter M$_{\mathrm{limit}}$. For satellite galaxies, the star formation is mostly reduced due to environmental quenching. Our corresponding parameter t$_{\mathrm{quench}}$ separates into recently and long-since merged subhalos, hosting blue and red satellite galaxies, respectively.\\
The properties matched in the abundance matching are the halo or subhalo mass and the galaxy absolute magnitude, corresponding to the luminosity. The most luminous galaxies reside in the most massive halos, filling up the halos and subhalos from massive to light. Our model performs this matching in narrow redshift bins, leading to an accurate, redshift dependent relation between mass and luminosity.\\
We showed the comparison of our simulations with data from the public Dark Energy Survey Year 1 (DES Y1). To ensure completeness of our simulations at very low redshifts, we have restricted the comparison to a magnitude range of $i<20$, leaving only bright galaxies. In section \ref{sec:res_dist_func}, we showed that the histograms of apparent magnitudes and galaxy distribution in colour space are in agreement with the data.\\
The results of our comparison for the clustering were shown in section \ref{sec:res_clustering}. We can only trust the comparison between data and our simulations in the the 2-halo regime, due to the approximate nature of \texttt{PINOCCHIO} and the subhalo model introduced by \cite{2021_berner}. We have calculated the magnitude-dependent clustering by binning in the $i$-band. Furthermore, we looked at the colour-dependent clustering by splitting the catalogue into a redder and a bluer sample, according to a cut in colour space. On reliable angular scales, the correlation functions of the simulations and the data are in statistical agreement.\\
By comparing to the data, we constrained the two SHAM parameters purely based on clustering. We found that good values within the best fit contour are M$_{\mathrm{limit}} =$ $8 \cdot 10^{12}$ h$^{-1}$M$_{\mathrm{\odot}}$ and t$_{\mathrm{quench}} =$ 2.0 Gyr. There is a degeneracy between the two parameters, since increasing either parameter increases the clustering of the galaxies. We could not reduce this degeneracy by distinguishing between red and blue galaxies, since the effect of changing the parameters along the degeneracy is only measurable in the 1-halo regime, where we cannot compare data and simulations reliably.\\
Compared to models matching galaxies to halos based on a full N-body simulation or even compared to a hydrodynamical simulation, our simulations take little computational effort. Running the \texttt{PINOCCHIO} lightcone (once) took 19 minutes on 1032 CPU cores of the Euler Cluster from ETH Zurich\footnote{https://scicomp.ethz.ch/wiki/Euler}, with 2048$^3$ simulation particles. Creating the subhalo catalogue based on the merger history takes about 10 hours, on one CPU core. The SHAM simulation itself is much faster and can be run multiple times for one halo/subhalo ligthcone. Sampling from the luminosity functions before the SHAM takes about 4 minutes, the SHAM model itself until the final galaxy catalogue about 28 minutes, both on one CPU. The SHAM model requires about 30 GB of RAM. The reported values are for the simulations of DES Y1 for the redshift range z<0.75.\\
Future work may include an extension to fainter galaxies and higher redshifts, and applications to current wide surveys. Further projects in progress include the development of a bias model based on the presented SHAM model, for the galaxies in \texttt{UCat} and \texttt{UFig}. Additionally, the effect of clustering on weak lensing systematics can be investigated using our framework. Matching the initial conditions and therefore the cosmic seed of our halo-subhalo lightcone to a full N-body particle simulation may open doors for further applications, including probe combinations. 

\acknowledgments
We thank J\"org Herbel at ETH Zurich for his helpful comments on \texttt{UFig}. We further thank Adam Amara at ETH Zurich, University of Portsmouth and University of Surrey for his input on running simulations in previous work. We would like to thank Pascal Hitz and Silvan Fischbacher, both at ETH Zurich, for valuable discussions on simulations and numerical issues. We thank Martina Fagioli at ETH Zurich for helpful discussions on galaxy populations. This project was supported in part by SNSF grant 200021\_192243.\\

\noindent This project used public archival data from the Dark Energy Survey (DES). Funding for the DES Projects has been provided by the U.S. Department of Energy, the U.S. National Science Foundation, the Ministry of Science and Education of Spain, the Science and Technology FacilitiesCouncil of the United Kingdom, the Higher Education Funding Council for England, the National Center for Supercomputing Applications at the University of Illinois at Urbana-Champaign, the Kavli Institute of Cosmological Physics at the University of Chicago, the Center for Cosmology and Astro-Particle Physics at the Ohio State University, the Mitchell Institute for Fundamental Physics and Astronomy at Texas A\&M University, Financiadora de Estudos e Projetos, Funda{\c c}{\~a}o Carlos Chagas Filho de Amparo {\`a} Pesquisa do Estado do Rio de Janeiro, Conselho Nacional de Desenvolvimento Cient{\'i}fico e Tecnol{\'o}gico and the Minist{\'e}rio da Ci{\^e}ncia, Tecnologia e Inova{\c c}{\~a}o, the Deutsche Forschungsgemeinschaft, and the Collaborating Institutions in the Dark Energy Survey.\\
The Collaborating Institutions are Argonne National Laboratory, the University of California at Santa Cruz, the University of Cambridge, Centro de Investigaciones Energ{\'e}ticas, Medioambientales y Tecnol{\'o}gicas-Madrid, the University of Chicago, University College London, the DES-Brazil Consortium, the University of Edinburgh, the Eidgen{\"o}ssische Technische Hochschule (ETH) Z{\"u}rich,  Fermi National Accelerator Laboratory, the University of Illinois at Urbana-Champaign, the Institut de Ci{\`e}ncies de l'Espai (IEEC/CSIC), the Institut de F{\'i}sica d'Altes Energies, Lawrence Berkeley National Laboratory, the Ludwig-Maximilians Universit{\"a}t M{\"u}nchen and the associated Excellence Cluster Universe, the University of Michigan, the National Optical Astronomy Observatory, the University of Nottingham, The Ohio State University, the OzDES Membership Consortium, the University of Pennsylvania, the University of Portsmouth, SLAC National Accelerator Laboratory, Stanford University, the University of Sussex, and Texas A\&M University.\\
Based in part on observations at Cerro Tololo Inter-American Observatory, National Optical Astronomy Observatory, which is operated by the Association of Universities for Research in Astronomy (AURA) under a cooperative agreement with the National Science Foundation.
\newpage 
\appendix
\section{Luminosity function of red and blue galaxies}
\label{sec:app_a}
\begin{figure}[b!]
\centering 
\includegraphics[width=0.8\textwidth,angle=0]{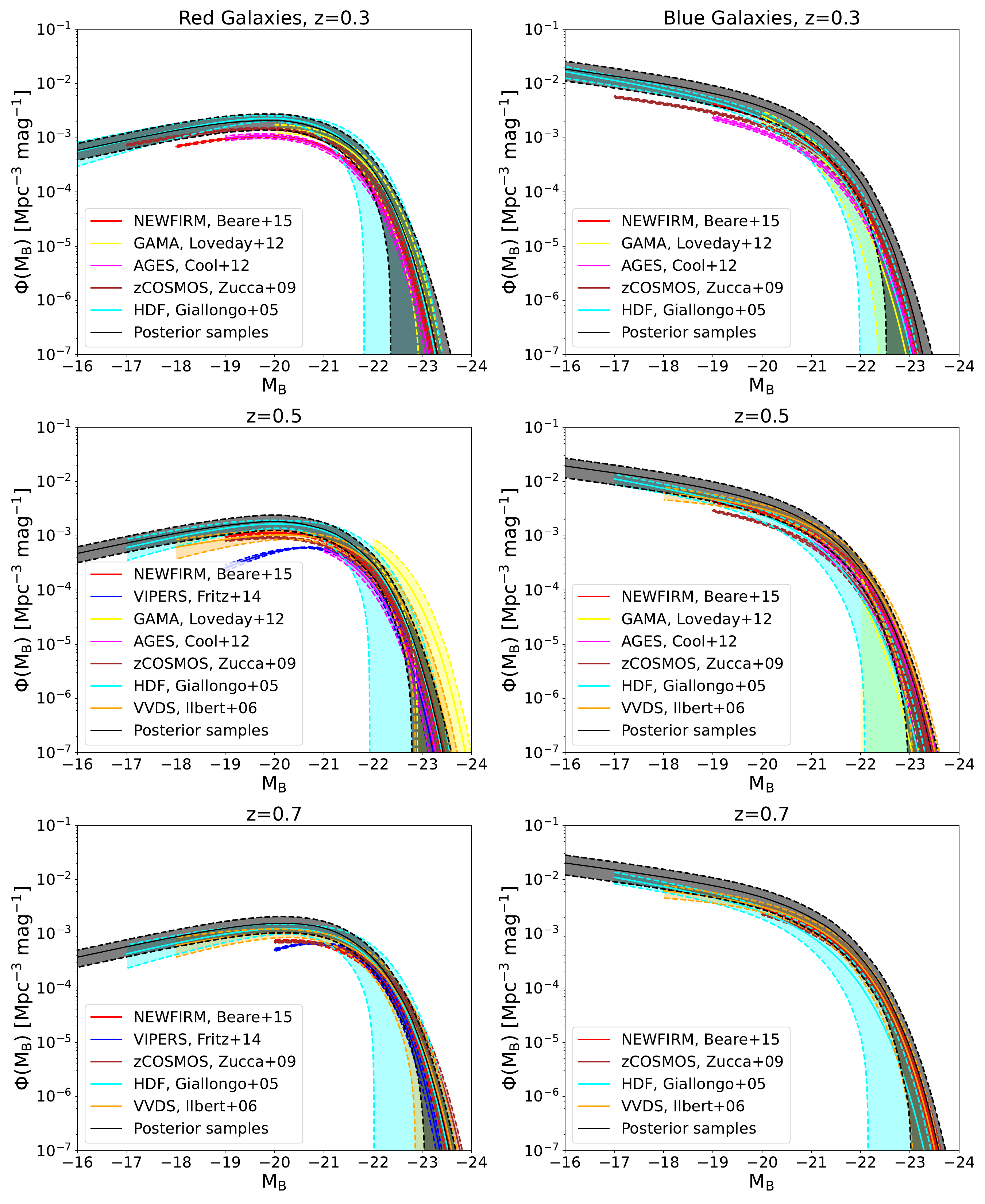}
\caption{The luminosity functions from the posterior distributions from \cite{Herbel_2017} (in black) for red (left) and blue galaxies (right) compared to literature luminosity functions. The functions from literature are from the works of \cite{2015_beare, 2014_fritz, 2012_loveday, 2012_cool, 2009_zucca, 2005_giallongo, 2006_ilbert} (in red, blue, yellow, magenta, brown, cyan and orange, respectively). For the posterior samples, the median and 1$\sigma$ uncertainties are shown. For the literature luminosity functions, the bands correspond to 1$\sigma$ errors. The redshifts chosen are z=0.3 (top), z=0.5 (middle) and z=0.7 (bottom).}
\label{fig:lf_posterior_literature}
\end{figure}
\noindent In this appendix, we show a comparison between the red and blue luminosity functions used for the galaxy clustering simulations in this work and observed luminosity functions from other works. The comparison is displayed in figure \ref{fig:lf_posterior_literature}, similar to figures 8 and 9 in \cite{Tortorelli_2020}, though with the posterior distribution from the ABC in \cite{Herbel_2017} used for this work. The literature studies shown in the comparison are from \cite{2015_beare, 2014_fritz, 2012_loveday, 2012_cool, 2009_zucca, 2005_giallongo, 2006_ilbert}, same as in the analysis presented in \cite{Tortorelli_2020}. \\
The redshifts chosen for this figure are z=0.3, z=0.5 and z=0.7. We do not show a comparison at lower redshifts, since the available literature studies do not span that redshift range. The luminosity functions used here are in good agreement with the literature, both for blue and red galaxies, for each considered redshift. The constraining power of the posterior from \cite{Herbel_2017} on the luminosity functions is similar to that from the different literature studies.

\bibliographystyle{ieeetr}
\bibliography{refs}

\end{document}